\documentstyle[12pt]{article}
\begin{document}
\begin{large}

\leftline{\hskip 12 truecm IJS.TP.99/31} 
\vspace{5mm}

\centerline{\bf UNIFICATION of SPINS AND CHARGES }
\centerline{\bf in GRASSMANN SPACE and}
\centerline{\bf in  SPACE of DIFFERENTIAL FORMS}

\end{large}

\vspace{0.8cm}
\centerline{\rm NORMA MANKO\v C BOR\v STNIK\footnote[1]{The
invited talk, presented on the International workshop
Bled99/BSM, "What comes beyond the Standard model" and on the
International Conference on Clifford 
Algebra and their Application in Mathematical Physics, Ixtapa,
27 June-4 July, 1999, Mexico, 
{\it Dept. of Physics, University of
Ljubljana, Jadranska 19, and }
{\it J. Stefan Institute, Jamova 39,
Ljubljana, 1111, Slovenia, and Primorska Institute of Natural
Sciences and Technology, C. Mare\v zganskega upora 2, 6000
Koper} {\it E-mail: norma.s.mankoc@ijs.si }}}

\vspace{2mm}
\centerline{\rm and}
\centerline{\rm HOLGER BECH NIELSEN}
\baselineskip=13pt
\centerline{\it Dept. of Physics,  Niels Bohr Institute,}
\centerline{\it Blegdamsvej 17, Copenhagen, DK-2100 and}
\centerline{\it Theoretical Physics Division, CERN, CH-1211
Geneva 23}
\centerline{\it E-mail: hbech@nbvims.nbi.dk}

\abstract{ Polynomials in Grassmann space can be used to
describe all the internal degrees of freedom of spinors, scalars
and vectors, that is their spins
and charges\cite{mankoc(92-97)}${ }$\cite{mankoc(99)}. It was
shown\cite{mankoc(99)} that K\" ahler spinors\cite{kahler(62)}, 
which are polynomials of differential forms, can be generalized
to describe not 
only spins of spinors but also spins of vectors as well as spins
and charges of scalars, vectors and spinors.
If the space (ordinary and noncommutative) has 14 dimensions or
more, the appropriate spontaneous break of 
symmetry leads gravity in $d$ dimensions to manifest in four
dimensional subspace as ordinary gravity and all needed gauge
fields as well as the Yukawa couplings. Both approaches, the K\"
ahler's one (if generalized) and our, manifest
four generations of massless fermions, which are 
left handed $SU(2)$ doublets and right handed $SU(2)$ singlets.
In this talk a possible way of spontaneously broken symmetries
is pointed out on the level of canonical momentum.
}

\section{  Introduction. \label{intr}}

Our world has besides the ordinary space-time the internal space
of spins and charges. Without the internal space, no matter
would exist and accordingly no complexity, which is  needed for the life to
exist. We have 
shown\cite{mankoc(92-97)} how a space of anticommuting
coordinates can be used to describe 
spins and charges of not only fermions but also of bosons,
unifying spins and charges for either fermions or for bosons and
that 
gravity in d dimensions manifests after appropriate break of
symmetry in $d=4$ dimensional subspace as ordinary gravity and
all  known gauge fields. 
K\" ahler\cite{kahler(62)} has shown how to use differential forms to
describe the spin of fermions. 
In the present talk we point out the analogy and nice relations
between the two different  ways of achieving the  appearance of 
spin one half degrees of freedom when starting from 
pure vectors and tensors. We comment the necessity of appearance
of four copies of Dirac fermions in both
approaches. This work was done together with H. B.
Nielsen\cite{mankoc(99)}. 
Comparing carefully the two approaches we generalize the K\"
ahler approach to describe also integer spins as well as charges
for either spinors or vectors, unifying spins and charges. 
\\
We present the possible Lagrange function for a free particle
and the canonical quantization of anticommuting
coordinates\cite{mankoc(92-97)}. 
Introducing vielbeins and spin connections,
we demonstrate how the spontaneous break of symmetry may lead to
the symmetries of the Standard model\cite{mankoc(92-97)}. In this
part of the talk (it has  been done together with 
A. Bor\v stnik\cite{mankocbor(99)}), we follow, how the break of
symmetries  from $SO(1,13)$ to
symmetries of the Standard model manifests on canonical
momentum.  
We show how the symmetry of the group $SO(1,13)$ breaks to
$SO(1,7)$ (leading to multiplets with left handed $SU(2)$
doublets and right handed $SU(2)$ singlets) and $SO(6)$, which
then leads to the $SO(1,3) \times SU(2) \times U(1) \times SU(3)
\times U(1)$. The two $U(1)$ symmetries enable besides the
hypercharge, needed in the Standard model, additional
hypercharge, which is nonzero for right handed $SU(2)$ singlet neutrino.  
For the pedagogical reasons we comment the break of
symmetry on the canonical momentum 
for the Standard model, that is from $SU(2) \times U(1)$ to
$U(1)$ as well.

\section{Dirac equations in Grassmann space.\label{deg}}

What we call quantum mechanics in Grassmann space\cite{mankoc(92-97)}
is the model for going beyond the Standard Model with extra
dimensions of ordinary and anticommuting coordinates,
describing spins and charges of either fermions or bosons in an
unique way. \\
In a $d$-dimensional space-time the internal degrees of freedom
of either spinors or vectors and scalars come from the odd
Grassmannian variables $\theta^a, \quad a \in \{ 0,1,2,3,5,
\cdot ,d \}$.  \\
We write wave functions describing either spinors or vectors 
in  the form
\begin{equation}
<\theta^a|\Phi>  = \sum_{i=0,1,..,3,5,..,d} \quad \sum_{\{ a_1<
a_2<...<a_i\}\in \{0,1,..,3,5,..,d\} }
\alpha_{a_1, a_2,...,a_i}\theta^{a_1} 
\theta^{a_2} \cdots \theta^{a_i},
\label{phi}
\end{equation} 
where the coefficients $ \alpha_{a_1, a_2,...,a_i}$ depend on
 commuting coordinates $ x^a, \; a \in \{0,1,2,3,5,..,d \}. $
The wave function space spanned over Grassmannian coordinate
space has the dimension $2^d$. Completely analogously to usual
quantum mechanics we have the operator for the conjugate
variable $\theta^a$ to be 
\begin{equation}
p^{\theta}_a = -i\overrightarrow{\partial}_a.
\label{pt}
\end{equation} 
The right arrow tells, that the derivation has to be performed
from the left hand side. These operators then obey the odd
Heisenberg algebra, which written by means of the generalized
commutators 
\begin{equation}
\{ A, B \}: = AB - ( -1)^{n_{AB}} BA,
\label{gc}
\end{equation}
where 
\begin{equation}
n_{AB}=\left\{ \begin{array}{rl} +1, \quad \hbox{if A and B have
Grassmann odd character}\\ 0, \quad \hbox{otherwise,}\end{array} \right.
\nonumber
\end{equation}   
takes the form
\begin{equation}
\{p^{\theta a},p^{\theta b}\} = 0 = \{\theta^a, \theta^b \} , \quad
\{ p^{\theta a}, \theta^b \} = -i \eta^{ab}.
\label{pptt}
\end{equation}
Here $\eta^{ab}$ is the flat metric $\eta = diag\{1,-1,-1,...\}$.
\\
We may  define the operators 
\begin{equation}
\tilde{a}^a := i(p^{\theta a}-i\theta^a), \quad\tilde{
\tilde{a}}{ }^a :=
-(p^{\theta a}+i\theta^a),
\label{at}
\end{equation}
for which we can show that the $\tilde{a}^a$'s among themselves 
fulfill the Clifford 
algebra as do also the $\tilde{\tilde{a}}{ }^a$'s, while they
mutually anticommute:
\begin{equation}
\{ \tilde{a}^a, \tilde{a}^b\} = 2\eta^{ab} = \{\tilde{\tilde{a}}{
}^a, 
\tilde{\tilde{a}}{ }^b\}, \quad \{\tilde
{a}^a, \tilde{\tilde{a}}{ }^b\} = 0.
\label{ateta}
\end{equation}

\noindent
We could recognize formally
\begin{equation}
{\rm either} \quad \tilde{a}^a p_a|\Phi> = 0, \qquad {\rm or} \quad 
\tilde{\tilde{a}}{ }^a p_a|\Phi> = 0 
\label{d}
\end{equation}
as the Dirac-like equation, because of the 
above generalized 
commutation relations.
Applying either the operator $\tilde{a}^a p_a$ or $\tilde{\tilde{a}}{
}^a p_a$ on the two equations we get the Klein-Gordon equation
$p^ap_a|\Phi> = 0$, where 
we define  $p_a = i\frac{\partial}{\partial x^a}$.
\\
One can check that none of the two equations
(\ref{d}) have solutions which would transform as spinors with
respect to the  generators of the Lorentz transformations,
when taken in analogy with the generators of the Lorentz
transformations in ordinary space ($L^{ab} = x^a p^b - x^b p^a $)
\begin{equation}
S^{ab}:= \theta^a p^{\theta b} - \theta^b p^{\theta a}.
\label{vecs}
\end{equation}
But we can write these generators as the sum
\begin{equation}
S^{ab} = \tilde{S}^{ab} + \tilde{\tilde{S}}{ }^{ab},
\quad \tilde{S}^{ab} := -\frac{i}{4} [\tilde{a}^a, \tilde{a}^b], \quad
\tilde{\tilde{S}}{ }^{ab} := -\frac{i}{4}[\tilde{\tilde{a}}{
}^a,\tilde{\tilde{a}}{ }^b],  
\label{vecsp}
\end{equation}
with $[A,B]:=AB-BA$ and recognize that the solutions of the two
equations (\ref{d}) now transform as spinors with respect to
either $ \tilde{S}^{ab}$ or $ \tilde{\tilde{S}}{ }^{ab}.$
\\
One also can easily see that the untilded, the single tilded and
the double tilded  
$S^{ab}$ obey the $d$-dimensional Lorentz generator algebra
\begin{equation}
\{ M^{ab}, M^{cd} \} = -i(M^{ad} \eta^{bc} + M^{bc} \eta^{ad} 
- M^{ac} \eta^{bd} - M^{bd} \eta^{ac}),
\label{lta}
\end{equation}
 when inserted for
$M^{ab}$. 
\\
We shall present our approach in more details in
section \ref{pbta} when pointing out the similarities between this approach
and the K\" ahler approach and generalizing the K\" ahler
approach.  In section \ref{lfmp} we shall present the Lagrange
function, which leads after canonical quantization in both
spaces, the ordinary one and the space of anticommuting
coordinates, to operators and equations presented in this section.

\section{K\" ahler formulation of spinors. \label{kfs} }

K\" ahler formulated\cite{kahler(62)} spinors in terms of wave
functions which are superpositions of the p-forms in the $d=4$ -
dimensional 
space. The 0-forms are scalars, the 1-forms are defined as dual
vectors to the (local) tangent spaces, the higher p-forms are
defined as antisymmetrized Cartesian (exterior ($\wedge $))
products of the 
one-form spaces. A general linear combination of forms is then
written 
\begin{equation}
u=u_0 + u_1 +...+u_d, \quad 
u_p =  \sum_{ i_1<i_2...<i_p}  a_{i_1i_2...i_p}\;
dx^{i_1}\wedge dx^{i_2}\wedge dx^{i_3} \wedge \cdots \wedge
dx^{i_p}.
\label{form} 
\end{equation}

\noindent
 The exterior product has the 
property of making the product of a p-form and a q-form to be a
(p+q)-form, if a p-form and a q-form have no common
differentials. One can define  also the Clifford
product ($\vee$) among the forms. The Clifford product $dx^a
\vee$ on a p-form is 
either a $p + 1$ form, if a p-form does not include a one form
$dx^a$, or a $p-1$ form, if a one form $dx^a$ is included in a
p-form.  \\
K\" ahler found how the Dirac equation could be written in terms
of differential forms
\begin{equation}
-i \delta u = m \vee u, \quad {\rm with} \quad \delta u =
\sum_{i=1}^3 dx^i \vee \frac{\partial u}{\partial 
x^i} - dt \vee \frac{\partial u}{\partial t}.
\label{dk1}
\end{equation}
with $u$ defined in Eq.(\ref{form}). The symbol $\delta$ denotes
the inner differentiation, $ a \in \{ 0,1,2,3\}$ and $m$ means
the electron mass.
\\
For a free massless particle living in a d dimensional
space-time  Eq.(\ref{dk1}) can be rewritten in the form
\begin{equation}
dx^a \vee p_a \;\;u = 0, \quad a = 0,1,2,3,5,...,d.
\label{vee}
\end{equation}
The wave function describing the state of
the spin one half particle is packed into the exterior algebra 
function $u$.

\section{ Parallelism between the two approaches. \label{pbta}}

We demonstrate the parallelism between the K\" ahler\cite{kahler(62)}
and our\cite{mankoc(92-97)}  approach in 
steps, first paying attention on spin $\frac{1}{2}$ only, as K\"
ahler did. Using simple and transparent definitions of the exterior
and interior product in Grassmann space, we
generalize\cite{mankoc(99)} the K\" 
ahler approach first by defining the two kinds of $\delta$
(Eq.(\ref{dk1})) 
operators on the space of p-forms and accordingly three kinds of
the generators of the Lorentz transformations, two of the
spinorial and one of the vectorial character. We try to put
clearly 
forward how the spinorial degrees of freedom emerge out of
vector objects like the 1-forms or $\theta^a$'s. We then
generalize the p-forms to describe not only spins but also
charges of spin $ \frac{1}{2}$ and spin 0 and 1 objects,
unifying also in the space of forms spins and charges,
separately for fermions and separately for bosons. 

\subsection{  Dirac-K\" ahler equation and  Dirac equation in
Grassmann  space for massless particles.\label{dkdg}}

We present here, side by side, the operators in the space of
differential forms and in Grassmann space: the ''exterior''
product 
\begin{equation}
dx^a \; \wedge dx^b \; \wedge \cdots , \quad \theta^a \theta^b
\cdots, 
\label{ext}
\end{equation}
the operator of ''differentiation''
\begin{equation}
-i\; e^a, \quad p^{\theta a} = -i\; \overrightarrow{\partial}^a
= -i \frac{\overrightarrow{\partial}}{\partial \theta_a}, 
\label{ea}
\end{equation}
and the two superpositions  
\begin{eqnarray}
dx^a \;\tilde{\vee}: = dx^a \wedge + \; e^a,  \quad
\tilde{a}^a:= i\;(p^{\theta a} - i \theta^a),
\nonumber
\\
dx^a \;\tilde{\tilde{\vee}}: = i\;( dx^a \wedge - \; e^a),  \quad
\tilde{\tilde{a}}{ }^a:= -(p^{\theta a} + i \theta^a).
\label{att}
\end{eqnarray}
The superposition, which we signed by $\tilde{ }$ is the one used by
K\" ahler (Eqs.(\ref{dk1})).
\\
One easily finds (see Eqs.(\ref{at},\ref{ateta})) the commutation
relations, understood in the generalized sense of Eq.(\ref{gc})
\begin{eqnarray}
\{ dx^a \; \tilde{\vee},\; dx^b \; \tilde{\vee} \} = 2
\eta^{ab}, \qquad 
\{ \tilde{a}^a, \tilde{a}^b \}  = 2 \eta^{ab},
\nonumber
\\
\{ dx^a \; \tilde{\tilde{\vee}},\; dx^b \;\tilde{\tilde{\vee}} \} = 2
\eta^{ab}, \qquad 
\{ \tilde{\tilde{a}}{ }^a, \tilde{\tilde{a}}{ }^b \} = 2
\eta^{ab}. 
\label{ctilde}
\end{eqnarray}
Since $ \{ e^a , dx^b \; \wedge \} = \eta^{ab}$ and $ \{e^a , e^b
 \} = 0 = \{dx^a \; \wedge, dx^b \; \wedge
 \}$, while $\{ -i p^{\theta a}, 
\theta^b \} = \eta ^{ab}$ and $\{ p^{\theta a},  p^{\theta b} \}
= 0 = \{ \theta^a, \theta^b \}$, it is obvious that $ e^a 
 $ plays in the p-form formalism the role of the
derivative with respect to a differential $1-$form, similarly as
$ip^{\theta a}$ does with respect to a Grassmann coordinate. 
\\
We find for both approaches the  Dirac-like equations:
\begin{eqnarray}
dx^a\; \tilde{\vee}\; p_a\; u = 0,   \quad
\tilde{a}^a\; p_a\; \Phi(\theta^a) = 0,
\nonumber
\\
dx^a\; \tilde{\tilde{\vee}}\; p_a\; u = 0,  \quad \quad
\tilde{\tilde{a}}{ }^a\; p_a\; \Phi(\theta^a) = 0. 
\label{dkt}
\end{eqnarray}

\noindent 
Taking into account the above definitions it follows that
\begin{equation}
dx^a \; \tilde{ \vee}\; p_a\;\; dx^b\; \tilde{\vee}\;
p_b \;u = p^a\; p_a \; u = 0, 
\qquad \tilde{a}^a\; p_a \;\;
\tilde{a}^b\; p_b\; \Phi(\theta^b) = p^a \; p_a\; \Phi(\theta^b) = 0.
\end{equation}
We see that either $dx^a \; \tilde{ \vee}\; p_a \; u = 0\; $ or
$\; dx^a\; \tilde{\tilde{ \vee}} \; p_a\; u = 0 ,$ similarly as either
$\tilde{a}^a \;p_a \;\Phi(\theta^a) = 0\;
$ or $\; \tilde{\tilde{a}}{ }^a \; p_a \; \Phi(\theta^a) = 0 $
can represent the Dirac-like equation.
\\
Both, $dx^a \; \tilde{\vee}$ and $dx^a \; \tilde{\tilde{\vee}} $
define the algebra of the $\gamma^a$ matrices 
and so  do both $\tilde{a}^a $ and $ \tilde{\tilde{a}}{
}^a$. One would thus be tempted to identify
\begin{equation}
\gamma_{ \hbox{naive} }^a := dx^a \; \tilde{\vee},\qquad \hbox{ or } 
\qquad \gamma_{ \hbox{naive} }^a :=  \tilde{a}^a.
\label{naive}
\end{equation}
But there is a large freedom in defining what to identify with
the gamma-matrices, because except when using $\gamma^0$ as
a parity operation, one has an even number of gamma matrices 
occurring in the physical applications such as construction of 
currents $\bar{\psi}\gamma^a\psi$ or for the Lorentz generators 
on spinors $\frac{-i}{4}\; [\gamma^a,\gamma^b]$. Then all the
gamma matrices can be multiplied by some factor provided it does 
disturb neither their algebra nor their even products.
This freedom might be used to solve, what seems a problem:
\\
Having an odd Grassmann character,
neither $\tilde{a}^a$ nor
$\tilde{\tilde{a}}{ }^a$ and similarly neither $dx^a\;
\tilde{\vee}$ nor $ dx^a\; \tilde{\tilde{\vee}}$ should be
recognized as the Dirac 
$\gamma^a$ operators, since they would change, when operating on
polynomials of $\theta^a$ or on superpositions of p-form,  
objects of an odd Grassmann
character to objects of an even Grassmann character. One
would, however, expect - since  Grassmann odd fields 
second quantize to fermions, while Grassmann even fields 
second quantize to bosons -  that the $\gamma^a$ operators do not
change the Grassmann character of  wave functions so that
the canonical quantization of Grassmann odd fields then
automatically assures the anticommuting relations between the
operators of the fermionic fields.
\\
We may propose that accordingly 
\begin{equation}
{\rm either} \quad \tilde{\gamma}^a: = i \; dx^0 \;
\tilde{\tilde{\vee}} \; dx^a \; \tilde{\vee}, \quad {\rm or} \qquad
\tilde{\gamma}^a = i \; \tilde{\tilde{a}}{ }^0 \; \tilde{a}^a 
\label{gamat}
\end{equation}
are recognized as the Dirac $\gamma^a$ operators operating on
the space of $p$-forms or polynomials of $\theta^a$'s,
respectively, since they both 
have an even Grassmann character and they both fulfill the
Clifford algebra $
\{ \tilde{\gamma}^a, \tilde{\gamma}^b \} = 2 \eta^{ab}.$
( The role of $\tilde{ }$ and $\tilde{\tilde{ }}$ can in either
the K\" ahler case or the case of polynomials in Grassmann space, be
exchanged. )
\\
The two definitions of  gamma-matrices  ((\ref{gamat}),
(\ref{naive})) make only a difference when $\gamma^0$-matrix is
used alone. This $\gamma^0$-matrix has to 
simulate the parity reflection which is 
\begin{equation}
{\rm either} \quad \vec{dx}\rightarrow -\vec{dx}, \quad {\rm or}
\quad 
\vec{\theta}\rightarrow - \vec{\theta}.
\label{parity}
\end{equation}
The ''ugly'' gamma-matrix identifications (\ref{gamat}) indeed
perform this operation. 
\\
K\" ahler did not connect evenness and oddness of the forms with
the statistics. He used the ''naive'' gamma-matrix
identifications (\ref{naive}).
The same can be said for
the  Becher-Joos (\cite{becher(82)}) paper.  

\subsection{ Generators of  Lorentz transformations.\label{glt}}

We are presenting the generators of the Lorentz
transformations of spinors for both approaches
\begin{equation}
M^{ab} = L^{ab} + {\cal S}^{ab}, \qquad L^{ab} = x^a p^b - x^b p^a.
\end{equation}
The two approaches  differ in the definition of the generators
of the Lorentz transformations in the internal space 
${\cal S}^{ab}$.
While K\" ahler suggested the definition for spin $\frac{1}{2}$
particles 
\begin{equation}
{\cal S}^{ab} = dx^a \wedge dx^b, \quad {\cal S}^{ab} u = \frac{1}{2}
( (dx^a \wedge dx^b) \vee u - u \vee ( dx^a \wedge dx^b )),
\label{sk}
\end{equation} 
in the Grassmann case\cite{mankoc(92-97)} the two kinds of the
operators 
${\cal S}^{ab} $ for spinors can be defined, presented in
Eqs.(\ref{vecsp}), with the properties 
\begin{equation}
[\tilde{S}^{ab}, \tilde{a}^c] = i(\eta^{ac} \tilde{a}^b -
\eta^{bc} \tilde{a}^a), \;\;         
[\tilde{\tilde{S}}{ }^{ab}, \tilde{\tilde{a}}{ }^c] =
i(\eta^{ac} \tilde{\tilde{a}}{ }^b -
\eta^{bc} \tilde{\tilde{a}}{ }^a), \;\; 
[\tilde{S}^{ab}, \tilde{\tilde{a}}{ }^c] = 0 =
[\tilde{\tilde{S}}{ }^{ab}, \tilde{a}^c].  
\end{equation}
Following the approach in Grassmann space one can also in the
K\" ahler case define two kinds of the Lorentz generators for
spinors, which (both) simplify  Eq.(\ref{sk})  
\begin{eqnarray}
\tilde{{\cal S}}^{ab} = -\frac{i}{4} [dx^a \; \wedge +
\;e^a, \; dx^b\; \wedge + \;
e^b ] ,
\quad \tilde{\tilde{\cal S}}{ }^{ab} =
\frac{i}{4} [dx^a\; \wedge -\; e^a 
,\; dx^b \; \wedge - \;e^b ],
\nonumber
\\
\tilde{{\cal S}}^{ab} = -\frac{i}{4} [\tilde{\gamma}^a,
\tilde{\gamma}^b].\qquad \qquad \qquad \qquad
\label{sabk}
\end{eqnarray}
The above definition enables us to define also in the K\" ahler
case 
the generators of the 
Lorentz transformations of the vectorial character 
\begin{equation}
{\cal S}^{ab} = \tilde{S}^{ab} + \tilde{\tilde{S}}{ }^{ab} = -i
( dx^a \wedge e^b  - dx^b \wedge e^a  ), \quad {\cal
S}^{ab} = \tilde{S}^{ab} + \tilde{\tilde{S}}{ }^{ab} = \theta^a 
p^{\theta b} - \theta^b p^{\theta a}.
\label{vecsk}
\end{equation}
 The operator
$ {\cal S} ^{ab} = -i (dx^a \wedge e^b  - dx^b \wedge e^a),  $ 
being applied on differential p-forms, transforms vectors
into vectors.

\subsection{Scalar product. \label{sptk}}

In our approach\cite{mankoc(92-97)} the scalar product between
the two functions $<\theta^a |\Phi_1>$  and $<\theta^a |\Phi_2>$ is
defined 
\begin{equation}
<\Phi_1|\Phi_2> = \int
d^d\;\theta\;\;(\omega\;<\theta^a|\Phi_1>)<\theta^a |\Phi_2> 
\label{scptk}
\end{equation}
and $\omega $ is a weight function
\begin{equation}
\omega = \prod_{i=0,1,..,d}\;\;(\theta^i +
\overrightarrow{\partial}^i), 
\nonumber
\end{equation}
which operates on only the first function $<\theta^a|\Phi_1>$
and 
\begin{equation}
\int \; d\theta^a = 0, \quad \int \; d^d\theta \theta^0
\theta^1...\theta^d = 1,\quad d^d\theta = \theta^d
...\theta^1 \theta^0.
\nonumber
\end{equation}
According to the above definition and Eq.(\ref{phi}) it follows
\begin{equation}
<\Phi^{(1)}|\Phi^{(2)}> = \sum_{0,d}\;\;
\sum_{\alpha_1<\alpha_2<..<\alpha_d}
\alpha^{(1)*}_{\alpha_1..\alpha_i}\;
\alpha^{(2)}_{\alpha_1..\alpha_i} 
\label{scpt}
\end{equation}
in complete analogy with the usual definition of the scalar
product in ordinary space. K\" ahler\cite{kahler(62)} defined 
the scalar product of two p-forms (Eq.(\ref{form})) as
\begin{equation}
<u^{(1)}|u^{(2)}> = \sum_{0,d}\;\;
\sum_{\alpha_1<\alpha_2<..<\alpha_d}
\alpha^{(1)*}_{\alpha_1..\alpha_i}\;
\alpha^{(2)}_{\alpha_1..\alpha_i}, 
\label{scpk}
\end{equation}
which agrees with Eq.(\ref{scpt}).

\subsection{ Four copies of Weyl bi-spinors  in  K\" ahler
or in approach in Grassmann space and vector
representations.\label{fws}}

 In the case of $d = 4$
one may arrange the space of $2^d $ vectors into four copies of 
two Weyl spinors, one left ( $<\tilde{\Gamma}^{(4)}> = - 1,\;
\;\; \Gamma^{(4)} = 
i\frac{(-2i)^2}{4!} \epsilon_{abcd} {\cal S}^{ab}{\cal S}^{cd}$
) and one right ( $< \tilde{\Gamma}^{(4)}> = 1$) handed (we
have made a choice of $\tilde{ }$ ),
in such a way that they are at the same time the eigen vectors
of the operators $\tilde{S}^{12}$ and the
$\tilde{S}^{03}$ and have  
either an odd or an even Grassmann character. These vectors
are in the K\" ahler approach the superpositions of p-forms and
in our\cite{mankoc(92-97)} approach the polynomials 
 of $\theta^m$'s, $m \in (0,1,2,3)$. The two Weyl
vectors of one copy of the Weyl bi-spinors are connected by the
$\tilde{\gamma}^m$ 
(Eq.(\ref{gamat})) operators, while the two copies of different
Grassmann character are connected by $\tilde{a}^a$ or $dx^a
\tilde{\vee}$, respectively. The two copies of an even
Grassmann character are connected by the ( a kind of a time
reversal operation) $ \theta^0 \; \rightarrow \; - \theta^0 $ or
equivalently $dx^0 \; \rightarrow \; - dx^0. $

We present in Table I four copies of the Weyl two spinors as
polynomials of $\theta^a$.
Replacing  $\theta^a$'s by $dx^a \wedge$ the
presentation for differential forms follow. Eigenstates are
orthonormalized according to the scalar product of Eq.(\ref{scpt})\\

\begin{center}
\begin{tabular}{||c|c||c||c|c|c|c|c||}
\hline
\hline
&&&&&&&\\ 
a & i &$<\theta|{}^a \Phi_i>$& $\tilde{S}^{12}$ & $
\tilde{S}^{03}$ & $\tilde{\Gamma}^{(4)}$& family & Grass. cha.\\ 
&&&&&&&\\
\hline
&&&&&&&\\
1 & 1 &$\frac{1}{2}(\tilde{a}^1 - i \tilde{a}^2)(\tilde{a}^0 -
\tilde{a}^3)$ &$ \frac{1}{2}$ &$ -\frac{i}{2}$&-1&&\\
1 & 2 &$-\frac{1}{2}(1 + i\tilde{a}^1 \tilde{a}^2)(1-
\tilde{a}^0 \tilde{a}^3) $ &$ -\frac{1}{2} $&$ \frac{i}{2}$&-1&&\\
&&&&&&I& even\\
2 & 1 &$\frac{1}{2}(\tilde{a}^1 - i \tilde{a}^2)(\tilde{a}^0 +
\tilde{a}^3)$ &$ \frac{1}{2}$ &$ \frac{i}{2}$&1&&\\
2 & 2 &$-\frac{1}{2}(1 + i\tilde{a}^1 \tilde{a}^2)(1+
\tilde{a}^0 \tilde{a}^3)$  &$ -\frac{1}{2} $&$ -\frac{i}{2}$&1&&\\
&&&&&&&\\
\hline
\hline
&&&&&&&\\
3 & 1 &$\frac{1}{2}(\tilde{a}^1 - i \tilde{a}^2)(1- \tilde{a}^0 
\tilde{a}^3)$ &$ \frac{1}{2}$ &$ \frac{i}{2}$&1&&\\
3 & 2 &$-\frac{1}{2}(1 + i\tilde{a}^1 \tilde{a}^2)(\tilde{a}^0 -
 \tilde{a}^3) $ &$ -\frac{1}{2} $&$ -\frac{i}{2}$&1&&\\
&&&&&&II& odd\\
4 & 1 &$\frac{1}{2}(\tilde{a}^1 - i \tilde{a}^2)(1 + \tilde{a}^0 
\tilde{a}^3)$ &$ \frac{1}{2}$ &$ -\frac{i}{2}$&-1&&\\
4 & 2 &$-\frac{1}{2}(1 + i\tilde{a}^1 \tilde{a}^2)(\tilde{a}^0 +
\tilde{a}^3)$  &$ -\frac{1}{2} $&$ \frac{i}{2}$&-1&&\\
&&&&&&&\\
\hline
\hline
&&&&&&&\\
5 & 1 &$\frac{1}{2}(1 -i \tilde{a}^1 \tilde{a}^2)(\tilde{a}^0 -
\tilde{a}^3)$ &$ \frac{1}{2}$ &$ -\frac{i}{2}$&-1&&\\
5 & 2 &$-\frac{1}{2}(\tilde{a}^1 +i \tilde{a}^2)(1-
\tilde{a}^0 \tilde{a}^3) $ &$ -\frac{1}{2} $&$ \frac{i}{2}$&-1&&\\
&&&&&&III&odd\\
6 & 1 &$\frac{1}{2}(1- i\tilde{a}^1 \tilde{a}^2)(\tilde{a}^0 +
\tilde{a}^3)$ &$ \frac{1}{2}$ &$  \frac{i}{2}$&1&&\\
6 & 2 &$-\frac{1}{2}(\tilde{a}^1 +i \tilde{a}^2)(1+
\tilde{a}^0 \tilde{a}^3)$  &$ -\frac{1}{2} $&$ -\frac{i}{2}$&1&&\\
&&&&&&&\\
\hline
\hline
&&&&&&&\\
7 & 1 &$\frac{1}{2}(1 -i\tilde{a}^1 \tilde{a}^2)(1- \tilde{a}^0 
\tilde{a}^3)$ &$ \frac{1}{2}$ &$ \frac{i}{2}$&1&&\\
7 & 2 &$-\frac{1}{2}(\tilde{a}^1 +i\tilde{a}^2)(\tilde{a}^0 -
 \tilde{a}^3) $ &$ -\frac{1}{2} $&$ -\frac{i}{2}$&1&&\\
&&&&&&IV&even\\
8 & 1 &$\frac{1}{2}(1-i\tilde{a}^1 \tilde{a}^2)(1 + \tilde{a}^0 
\tilde{a}^3)$ &$ \frac{1}{2}$ &$ -\frac{i}{2}$&-1&&\\
8 & 2 &$-\frac{1}{2}(\tilde{a}^1 +i \tilde{a}^2)(\tilde{a}^0 +
\tilde{a}^3)$  &$ -\frac{1}{2} $&$ \frac{i}{2}$&-1&&\\
&&&&&&&\\
\hline
\hline
\end{tabular}
\end{center}

\noindent
Table I: The polynomials of $\theta^m$, representing the four
times two Weyl spinors, are written. For each state the
eigenvalues of $\tilde{S}^{12}, \tilde{S}^{03},\;
\tilde{\Gamma}^{(4)}: = i \tilde{a}^0 \tilde{a}^1 \tilde{a}^2 \tilde{a}^3$ are
written. The Roman numerals tell the possible family number.
We use the relation $\tilde{a}^a |0> = \theta^a$.

Analyzing the irreducible representations of the group $SO(1,3)$
with respect to the generator of the Lorentz transformations of
the vectorial type\cite{mankoc(92-97)} (Eqs.( \ref{vecsk})) one finds
for d = 4 two scalars ( a scalar and 
a pseudo scalar), two three vectors (in the $SU(2) \times SU(2)
$ representation of $SO(1,3)$ denoted by $(1,0) $ and $(0,1)$
representation, respectively, with $<\Gamma^{(4)}> = \pm 1$) and
two four vectors. One can find the polynomial 
representation for this case in ref.\cite{mankoc(92-97)}.

\subsection{Generalization to extra dimensions.\label{ged}}

 It has  been
suggested\cite{mankoc(92-97)} that the Lorentz transformations in the
space of $ \theta^a$'s in  $d-4$ dimensions manifest
themselves as generators for charges observable  for
the four dimensional particles. Since both the
extra dimension spin degrees of freedom and the ordinary spin
degrees of freedom originate from the $\theta^a$'s or the forms
we have a unification of these internal degrees of freedom.
\\
Let us take as an example the model\cite{mankoc(92-97)} which has
$d=14$ and at first - at the high energy
level - $SO(1,13)$ Lorentz group, but which should be broken 
( in two steps ) to first $SO(1,7)\times SO(6)$ and then to
$SO(1,3)\times SU(3)\times SU(2) $. We shall comment on this
model in section \ref{bltsm}.

\section{ Appearance of spinors. \label{as}}

One of course  wonders about how it is at all possible 
that  the Dirac equation appears
for a {\em spinor} field  out of models with only scalar,
vector and tensor objects! 
 It  only can be done by  {\em exchanging} the Lorentz 
generators ${\cal S}^{ab}$ by the $\tilde{S}^{ab}$ say ( or the
$\tilde{\tilde{S}}{ }^{ab}$ if we 
choose them instead), see equations (\ref{vecsp}, \ref{sabk}).
This indeed means that one of the two kinds of operators
fulfilling the Clifford 
algebra and anticommuting with the other kind - it has been made a
choice of  $dx^a
\tilde{\tilde{\vee}} $ in the K\" ahler case and 
$\tilde{\tilde{a}}{ }^a $ in our approach  - are put
to zero in the operators of the Lorentz transformations; as well as in
all the operators representing physical quantities. The  use
of $dx^0 \tilde{\tilde{\vee}}$ or $ \tilde{\tilde{a}}{ }^0 $ in
the operator $\tilde{\gamma}^0$ is the exception,  only used to
simulate the Grassmann even  parity operation $\vec{dx}^a \to -\vec{dx}^a $
and $ \vec{\theta} \to - \vec{\theta}, $ respectively.
\\
We shall argue away (\cite{mankoc(92-97)}) the
$\tilde{\tilde{a}}{ }^a$'s in section \ref{lfmp}
on the ground of the action.

\section{Lagrange function for a free massless particle in
ordinary and Grassmann space and canonical quantization. \label{lfmp}}

We  present in this section the Lagrange function for a
particle which lives in a d-dimensional ordinary space
of commuting coordinates and in a d-dimensional Grassmann space
of anticommuting coordinates $ X^a \equiv \{ x^a, \theta^a \} $
and has its geodesics parametrized by an ordinary Grassmann even
parameter ($\tau$) and a Grassmann odd parameter($\xi$).  We
derive the Hamilton function and the corresponding Poisson
brackets and perform 
the canonical quantization, which leads to the Dirac equation with
operators presented in sections \ref{deg}, \ref{pbta}.

$ X^a = X^a(x^a,\theta^a,\tau,\xi)$ are called supercoordinates.
We define the dynamics of a particle by choosing the 
action \cite{mankoc(92-97),ikemori(87)}
$ I= \frac{1}{2} \int d \tau d \xi E E^i_A \partial_i X^a E^j_B
\partial_j X^b  \eta_{a b} \eta ^{A B},$
where $ \partial _i : = ( \partial _ \tau , {\overrightarrow
{\partial}} _\xi ), \tau^i = (\tau, \xi) $, while $ E^i _A
$ determines a metric on a two dimensional superspace $ \tau ^i
$ , $ E = det( E^i _A )$ . We choose $ \eta _{A A} = 0,   
\eta_{1 2} = 1 = \eta_{2 1} $, while $ \eta_{a b} $ is the
Minkowski metric with the diagonal elements $
(1,-1,-1,-1,$ $...,-1) $. The action is invariant under the Lorentz
transformations of supercoordinates: $X'{ }^a = \Lambda^a{ }_b X^b $.
Since a supermatrix $ E^i{ }_A $ transforms as a vector in a
two-dimensional superspace $\tau^i$ under general coordinate
transformations of $\tau^i$, $ E^i{ }_A \tau_i $ is invariant
under such transformations and so is $d^2 \tau E$. 
The action is locally supersymmetric.
 The inverse matrix $
E^A{ }_i$ 
is defined as follows: $E^i{ }_A E^B{ }_i = \delta^B{ }_A$.
 
Taking into account that either $ x^a $ or $ \theta^a $ depend
on an ordinary time parameter $ \tau $ and that $ \xi^2 = 0 $ ,
the geodesics can be described  as a
polynomial of  $ \xi $ as follows: 
$ X^a = x^a + \varepsilon \xi \theta^a $. We choose $
\varepsilon^2 $ 
to be equal either to $ +i $ or to $ -i $ so that it defines two
possible combinations of supercoordinates. Accordingly we
also choose  the metric  $ E^i { }_A $ : $ E^1{ }_1 = 1, E^1{
}_2 = - \varepsilon M, E^2{ }_1 = \xi, E^2{ }_2 = N -
\varepsilon \xi M $, with $ N $ and $ M $  Grassmann even and
odd parameters, respectively. We write $ \dot{A} =
\frac{d}{d\tau}A $, for any $ A $.

If we integrate the above action over the Grassmann odd
coordinate $d\xi$, the action for a superparticle follows:
\begin{equation}
 \int \; d\tau \;\;( \frac{1}{N} \dot{x}^a \dot{x}_a + \varepsilon^2
\dot{\theta}^a \theta_a - \frac{2\varepsilon^2 M}{N} \dot{x}^a
\theta_a ). 
\label{af}
\end{equation}
Defining the two momenta 
\begin{equation}
p^{\theta }_a : = \frac{ \overrightarrow{\partial} L}
{ \partial {\dot{\theta}^a}} = \epsilon^2 \theta^a,\;\; p_a : =
\frac{\partial L}{ \partial \dot{x}^a} = \frac{2}{N}( 
\dot{x}_a - M p^{\theta a}), 
\label{ptf}
\end{equation}
the two Euler-Lagrange equations follow:
\begin{equation} \frac{dp^a}{d \tau} = 0,\;\;\; \frac{dp^{\theta a}}{d \tau} =
\varepsilon ^2 \frac{M}{2} p^a. 
\label{el}
\end{equation}
Variation of the action (Eq.(\ref{af})) with respect to $M $ and
$N$ gives 
the two constraints
\begin{equation}
\chi^1: = p^a a^{\theta}_a = 0, \chi^2 = p^a p_a = 0, \;\;
a^{\theta}_a:= i p^{\theta}_a + \varepsilon^2 \theta_a,
\label{chi}
\end{equation} 
while  
$ \chi^3{ }_a: = - p^{\theta }_a + \epsilon^2 \theta_a = 0 $
(Eq.(\ref{ptf})) is the third type of constraints of the action(\ref{af}).
For $\varepsilon^2 = -i$ we find  that $
a^{\theta}{ }_a = \tilde{a} 
^a,\;\;$ which agrees with Eq.(\ref{at}), while $\chi^3{ }_a =
\tilde{\tilde a}_a = 0, $, which makes a choice between
$\tilde{a}^a$ and $\tilde{\tilde{a}}^a$. 

We find the generators  of the Lorentz transformations for the
action(\ref{af}) to be 
\begin{equation}
 M^{ a b} = L^{a b} + S^{a b} \;,\; L^{a b} = x^a p^b - x^b p^a
\;,\; S^{a b} = \theta^a p^{ \theta b} - \theta^b p^{ \theta a}
=  \tilde{S} ^{a b} + \tilde{\tilde{S}}{}^{a b},
\label{mabl}
\end{equation} 
which agree with definitions in Eq.(\ref{vecsp}) and 
show that parameters of the Lorentz transformations are the
same in both spaces.

We define the Hamilton function:
\begin{equation}
 H:= \dot{x}^a p_a + \dot{\theta}{ }^a p^{\theta}{ }_a - L =
\frac{1}{4} N p^a p_a + \frac{1}{2} M p^a (\tilde a_a +
i \tilde{\tilde a }{ }_a) 
\label{haml} 
\end{equation}
and the corresponding Poisson brackets
\begin{equation}
\{A,B\}_p=  
\frac{ \partial A}{ \partial x^a} \frac{ \partial B}{ \partial
p_a}  - \frac{ \partial A}{ \partial p_a} \frac{ \partial B}{ \partial
x^a} +  \frac{ \overrightarrow{ \partial A}}{\partial \theta
^a} \frac{ \overrightarrow{ \partial B}}{\partial p^\theta_a} +
  \frac{ \overrightarrow{ \partial A}}{\partial p^\theta_a} 
\frac{ \overrightarrow{ \partial B}}{\partial \theta^a}, 
\label{poil}
\end{equation} 
which fulfill  the algebra of the generalized
commutators\cite{mankoc(92-97)} of Eq.(\ref{gc}).

If we take into account the constraint $\chi^3{ }_a = \tilde{\tilde
a}{ }_a = 0\;$ in the Hamilton function (which just means that
instead of H the Hamilton function $ H + \sum_i \alpha^i
\chi^i + \sum_a \alpha^3{ }_a \chi^3{ }^a $ is taken, with parameters $
\alpha^i, i=1,2 $ and $ \alpha^3{ }_a = -\frac{M}{2} p_a,
a=0,1,2,3,5,..,d $ 
chosen on such a way that the Poisson brackets
of the three types of constraints with the new Hamilton function
are equal to zero) and in all dynamical quantities, we find:
\begin{equation}
H = \frac{1}{4} N p^a p_a + \frac{1}{2} M p^a \tilde a_a,\;\;
\chi^1 = p^a p_a = 0,\;\; \chi^2 = p^a \tilde a_a =
0,
\label{hamla}
\end{equation}
\begin{equation}
\dot{p}_a = \{ p_a, H \}_P = 0,\;\; \dot{\tilde a}{ }_a = \{ 
\tilde{a}_a, H \}_P = iM p_a,
\nonumber 
\end{equation}
which agrees with the Euler Lagrange equations (\ref{el}).

We further find 
\begin{equation}
 \dot {\chi}^i = \{ H, \chi^i \}_P = 0,\;\;i =1,2,\;\;\; \dot
{\chi}^3{ }_a = \{ H, \chi^3{ }_a \}_P = 0,\;\;a =
0,1,2,3,5,..,d,
\nonumber 
\end{equation}
which guarantees that the three
constraints will not change with the time parameter $\tau$ and
that $\dot{\tilde M}{ }^{ab} = 0 $, with $ \tilde M { }^{ab} =
L^{ab} + \tilde{S}^{ab}$,  saying that $ \tilde M{ }^{ab} $
is the constant of motion.

The Dirac brackets, which can be obtained from the Poisson
brackets of Eq.(\ref{poil}) by adding to these brackets on the right
hand side a term $ - \{A, \tilde{\tilde a}^c \}_P \cdot$ $ ( -
\frac{1}{2i} \eta_{ce} ) \cdot $ $ \{ \tilde{\tilde a}{ }^e, B
\}_P $, give  for the dynamical quantities, 
which are observables, the same results as the Poisson brackets.
This is true also for $ \tilde a^a,$ ( $\{ \tilde
a^a, \tilde a^b \}_D = i\eta^{ab} = \{ 
\tilde a^a, \tilde a^b \}_P $),  which is the
dynamical quantity but not  an observable since its odd
Grassmann character causes  supersymmetric
transformations. We also find that $\{ \tilde a^a, \tilde{\tilde
a}{ }^b \}_D 
= 0 = \{ \tilde a^a, \tilde{\tilde a}{ }^b \}_P $ .
The Dirac brackets give  different results only for the quantities
$ \theta^a $ and $ p^{\theta a} $ and  for $\tilde {\tilde
a}{ }^a $ among themselves: $ \{ \theta^a, p^{\theta b}
\}_P = \eta^{ab}, \{ \theta^a, p^{\theta b}
\}_D = \frac{1}{2} \eta^{ab} $, $ \{ \tilde {\tilde a}{ }^a, \tilde
{\tilde a}{ }^b \}_P = 2i \eta^{ab}, \{ \tilde {\tilde a}{ }^a,
\tilde {\tilde a}{ }^b \}_D = 0 $. According to the above  properties
of the Poisson brackets, we suggested\cite{mankoc(92-97)} that
in the quantization 
procedure the Poisson brackets (\ref{poil}) rather than the Dirac
brackets are used, so that variables $\tilde{\tilde a}^a $,
which are removed from all dynamical quantities, stay as
operators. Then $\tilde a^a $ and 
$\tilde{\tilde a}{ }^a $ are expressible with $\theta^a $ and
$p^{\theta a} $ (Eq.(\ref{at}))  and the algebra of linear operators
introduced in sections \ref{deg}, \ref{pbta} can be used. We
shall 
show, that  suggested quantization procedure leads to the
Dirac equation, which is the differential equation in ordinary
and Grassmann space and has all desired properties.

In the proposed quantization procedure $\; -i \{ A,B \}_p $ goes to
either a commutator or to an anticommutator, according to the
Poisson brackets (\ref{poil}). The operators $\theta ^a , p^{\theta a}
$ ( in the coordinate representation they become $ \theta^a 
\longrightarrow \theta^a , \; p^{\theta}_a \longrightarrow i 
\frac{\overrightarrow{\partial }}{\partial \theta^a} $) fulfill
the Grassmann odd Heisenberg algebra, while the operators 
$ \; \tilde{a}^a \; $ and $\; \tilde{\tilde{a}}{}^{a}\; $ fulfill
the Clifford algebra (Eq.(\ref{ateta})).

The constraints (Eqs.(\ref{chi})) lead to
the Weyl-like and  the Klein-Gordon equations
\begin{equation}
p^a \tilde{a} _a | \tilde{\Phi} > = 0 \;,\; p^a p_a |
\tilde{\Phi}> = 0 , \; {\rm with} \;  p^a \tilde{a}_a p^b \tilde{a}_b =
p^a p_a . 
\label{dkl}
\end{equation}
Trying to solve the eigenvalue problem $ \tilde{\tilde a}{ }^a 
| \tilde {\Phi} > = 0,\;\; a=(0,1,2,3,5,...,d), $ we find that no
solution of this eigenvalue problem exists, which means that
the third constraint $ \tilde{\tilde a}{ }^a = 0 $ can't be
fulfilled in the operator form (although we take it into account
in the operators for all dynamical variables in order that
operator equations would agree with classical equations). We can
only take it into account 
in the  expectation value form 
\begin{equation}
< \tilde{\Phi} | \tilde{\tilde a}{ }^a | \tilde{\Phi} > = 0.
\label{phittl}
\end{equation}
Since $ \tilde{\tilde a}{ }^a $ are Grassmann odd operators,
they change monomials (Eq.(\ref{phi})) of an Grassmann odd character
into monomials of an Grassmann even character and opposite,
which is the supersymmetry transformation.
It means that Eq.(\ref{phittl}) is fulfilled for monomials of either odd
or even Grassmann character and that superpositions of the
Grassmann odd and the Grassmann even monomials are not solutions
for this system. 

We  define the projectors
\begin{equation} 
 P_{\pm} = \frac{1}{2} ( 1 \pm  \sqrt{ (-)^{\tilde
\Upsilon \tilde{\tilde 
\Upsilon}}} \tilde{ \Upsilon} \tilde{\tilde \Upsilon}),\;\;\;\;
(P_{\pm})^2 = 
P_{\pm}, 
\label{proj} 
\end{equation}
where $\tilde \Upsilon$ and $ \tilde{\tilde \Upsilon}$ are the two
operators  defined  for any dimension d as follows
$ \tilde \Upsilon = \; i^{\alpha} $ $ \prod_{a=0,1,2,3,5,..,d} \tilde{
a}{ }^a $ $\sqrt{\eta^{aa}},$ $\;\; \tilde{\tilde \Upsilon} =
i^{\alpha} \prod_{a=0,1,2,3,5,..,d} 
$ $\tilde{ \tilde{a}}{ }^a \sqrt{\eta^{aa}},$
with $\alpha $ equal either to $ d/2 $ or to $ (d-1)/2 $ for
even and odd dimension $d$ of the space, respectively. 
It can be checked that $( \tilde \Upsilon )^2 = 1 = ( \tilde{
\tilde{\Upsilon}} )^2 $.

We can use the projector $P_{\pm}$ of Eq.(\ref{proj}) to project
out of monomials  either the Grassmann odd or the Grassmann even
part. Since this projector commutes with the Hamilton function $
 (\{ P_{\pm}, H \} = 0 ) $,  it means that eigenfunctions of $
H $, which fulfil the Eq.(\ref{phittl}), have either an odd or an even
Grassmann character. 
In order that in the second quantization procedure  fields
$ | \tilde{\Phi} > $ would describe fermions, it is meaningful
to accept  in the fermion case Grassmann  odd monomials only.
(See discussions in ref.(\cite{mankoc(99)}).)

\section{Particles in gauge fields. \label{pgf}}

The dynamics of a point particle in gauge fields, the
gravitational in $d$-dimensions, which then, as we shall show,
manifests in the subspace $d=4$ as ordinary gravity and all the
Yang-Mills fields, can be obtained by 
transforming  vectors from a freely falling to an external 
coordinate system \cite{wess(83)}.  
To do this, supervielbeins ${\bf
e}^{a}{ } _{\mu} $ have to be 
introduced, which in our case depend on ordinary and on
Grassmann coordinates, as well as on 
two types of parameters $ \tau^i = ( \tau, \xi ) $.  The index a 
refers to a freely falling coordinate system ( a Lorentz index),
the index $\mu$ refers to an external coordinate system ( an
Einstein index). 
 
We write the transformation of vectors as follows
$ \partial_i X^a= {\bf e}^{ a} { }_{\mu} \partial_i X^{\mu} \;,\;
\partial_i X^{\mu} = {\bf f}^{ \mu} { }_a \partial_i X^a \;,\;
\partial_i $ $= ( \partial_{\tau} , \partial_{\xi} ).$
From here it follows that
$ {\bf e}^{ a} { }_{\mu} {\bf f}^{ \mu} { }_b = \delta^a { }_b \;,\;  
{\bf f}^{ \mu} { }_{a} {\bf e}^{ a} { }_{\nu} = \delta^{\mu} {
}_{\nu}.$ 

Again we make a Taylor expansion of vielbeins with respect to 
$ \xi,\; $ 
$ {\bf e}^{ a} { }_{\mu} = e^{a} { }_{\mu} + \varepsilon^2 \xi
\theta^b e^{a} { }_{ \mu b} \;,\; {\bf f}^{ \mu} { }_a $ $= f^{
\mu} { }_a - \varepsilon^2 \xi \theta^b
f^{\mu} { }_{a b}.$

Both expansion coefficients  again depend  on ordinary
and on Grassmann coordinates. Having an even Grassmann character
$ e^{a} { }_{\mu}$  will describe the spin 2 part of a
gravitational field. The coefficients $ 
e^{a} { }_{\mu b}$  define the
spin connections\cite{mankoc(92-97)}. 

It follows that
$   e^{ a} { }_{\mu} f^{\mu} { }_b = \delta^a { }_b \;,\;  
f^{\mu} { }_{a} e^{a} { }_{\nu} = \delta^{\mu} { }_{\nu}
\;,\; e^{a} { }_{\mu b} f^{\mu} { }_c = e^{a} { }_{\mu}
f^{ \mu} { }_{c b}.$

We find the metric tensor ${\bf g}_{\mu \nu} = {\bf e}^{a}
{ }_{\mu} {\bf e}^{ }_{a \nu} ,\;
{\bf g}^{ \mu \nu} ={\bf f}^{ \mu} { }_{a} {\bf f}^{\nu a}$. 

Rewriting the action from section \ref{lfmp} in terms of an external
coordinate system, using the Taylor expansion of 
supercoordinates $ X^{\mu}$ and superfields $ {\bf{e}}^{a} {
}_{\mu}$ and
integrating the action over the Grassmann odd parameter $\xi$,
the action follows
\begin{eqnarray}
 I=\int d\tau \; \{ \frac{1}{N} g^{\mu \nu} \dot{x}^\mu
\dot{x}^\nu \; - \; \varepsilon^2 \frac{ 2 M}{N} \theta_a e^{a} {
}_{\mu} \dot{x}^\mu \; + \; \varepsilon^2 \frac{1}{2}(
\dot{\theta}^\mu \theta_a -\theta_a \dot{\theta}^\mu) e^{a} {
}_{\mu} \; + 
\nonumber\\ 
 + \;  \varepsilon^2 \frac{1}{2} (\theta^b \theta_a
-\theta_{a} \theta^b ) e^{a} { }_{  \mu b} \dot{x}^\mu \} ,
\label{acgr}
\end{eqnarray}
which defines the two momenta of the system
$ p_{\mu} = \frac{\partial L}{\partial \dot{x}^\mu} = p_{0 \mu} +
 \frac{1}{2} \tilde{S}^{ab} e_{a \mu b} , \;\;
 p^\theta_\mu = -i \theta_a e^{a} { }_{\mu}
$
( $\varepsilon^2 = -i $ ).
Here $ p_{0 \mu} $ are the canonical (covariant) momenta of a
particle. 
For $ p^{\theta}_{a} = p^{\theta}_{\mu} f^{\mu} { }_{a}$, it follows
that $ p^{\theta}_{a}$ is proportional to $\theta_{a}$. Then $
\tilde{a}_{a} = i 
(p^{\theta}_{a} - i \theta_{a}), 
$ while $ \tilde{\tilde{a}}_{a}= 0 $.  We may further write
\begin{equation}
 p_{ 0 \mu} = p_{ \mu} - \frac{1}{2} \tilde{S}^{a b} e_{a \mu b}
= p_{ \mu} - \frac{1}{2} \tilde{S}^{a b} \omega_{a b \mu} \;,\;
\omega_{a b \mu}=\frac{1}{2} (e_{a \mu b} - e_{b \mu a}),
\label{cmg}
\end{equation} 
which is the usual expression for the covariant momenta in
gauge gravitational fields\cite{wess(83)}.
One can find  the two constraints
\begin{equation}
 p_0^\mu p_{0 \mu} = 0 = p_{0 \mu} f^{\mu} { }_a \tilde{a}^a .
\label{cong} 
\end{equation}

We shall comment on the break of symmetries which leads in $d=4$
dimensional subspace as ordinary gravity and all the gauge field
in section \ref{bltsm}.

\section{ Breaking 
 $SO(1,13)$ through $SO(1,7) \times SO(6)$ to 
$SO(1,3) \times SU(2) \times U(1) \times SU(3)$. \label{bltsm}}

In this section, we shall first discuss a possible break of
symmetry, which leads from the unified theory of only spins and
gravity in d dimensions to spins and charges and to 
 the symmetries and assumptions of the Standard model, on the
algebraic level (\ref{ac}).
We shall then comment on the break of symmetries on the level of
canonical momentum, first for the Standard model case (\ref{smodc}), to
only demonstrate the way of the break, and then for the general
case, that is 
for the particle in the presence of the gravitational field
(\ref{dar}). 

We shall present as well the possible explanation for that
postulate of the Standard model, which requires that only left
handed weak charged massless doublets and right handed weak
charged massless singlets exist, and accordingly   
 connect spins and charges of fermions.

\subsection{Algebraic considerations of symmetries. \label{ac}}

 The algebra of the group $ SO(1,d-1) $ {\it or} $ SO(d) $ 
contains \cite{mankoc(92-97)}  $ n $ subalgebras defined by 
operators $ \tau ^{A i}, A = 1,n ; i = 1,n_A $,
where $ n_A $ is the number of elements of each
subalgebra, with the properties 
\begin{equation}
 [ \tau ^{Ai} , \tau ^{B j} ] = i \delta ^{AB
} f^{A ijk } \tau ^{A k}, 
\label{taug}
\end{equation}
if operators $ \tau ^{A i} $  can be expressed as
linear superpositions  of operators $ M^{ab} $ 
\begin{equation}
 \tau^{A i} = {\it c} ^{A i} { }_{ab} M^{ab}, \;\;
{\it c} ^{A i}{ }_{ab} = - {\it c} ^{A i}{ }_{ba}, \;\;
A=1,n, \;\;
i=1,n_{A}, \;\;a,b=1,d. 
\label{tauga} 
\end{equation}

Here $ f^{A ijk} $ are structure constants
of the ($ A $) subgroup with $ n_{A} $ operators.
According to the three kinds of operators $ {\cal S}^{ab} $, two of
spinorial and one of vectorial character, there are  three kinds
of operators $ \tau^{A i} $ defining subalgebras of
spinorial and vectorial character, respectively, those of
spinorial types being expressed with either $ \tilde S^{ab} $ or
$ \tilde{ \tilde S}{ }^{ab} $ and those of vectorial type being
expressed by $ S^{ab} $. All three kinds
of operators are, according to Eq.(\ref{taug}), 
defined by the same coefficients $ {\it c}^{A i} { }_{ab} $
and the same structure constants $ f^{A i j k } $.
From Eq.(\ref{taug}) the following relations among constants ${\it
c}^{A i}{ }_{ab} $ follow
\begin{equation}
 -4 {\it c}^{A i}{ }_{ab} {\it c}^{B j b}{ }_c -
\delta^{A B} f^{A ijk} {\it c}^{A k}{ }_{ac}
= 0. 
\label{taugb}
\end{equation}

When we look for coefficients $ c^{A i}{ }_{ab} $ which
express operators $ \tau ^{A i} $, forming a subalgebra
$ SU(n) $ of an algebra $ SO(2n) $ in terms of $ M^{ab} $, the
procedure is rather simple \cite{georgi(82),mankoc(92-97)}. We
find: 
\begin{equation} 
 \tau^{A m} = -\frac{i}{2} (\tilde \sigma^{A m})_{jk}
 \{ M^{(2j-1) (2k-1)} +
M^{(2j) (2k)} + i M^{(2j) (2k-1)} - i M^{(2j-1) (2k)}
\}.
\label{tausab} 
\end{equation}
Here $(\tilde \sigma^{A m})_{jk}$ are the traceless matrices
which form the algebra of $ SU(n) $.
One can easily prove that operators $ \tau^{A m} $ fulfill the
algebra of the group $ SU(n) $  for any of three
choices for operators $ M^{ab} : S^{ab}, \tilde S^{ab},
\tilde{\tilde S}{ }^{ab}$.

While the coefficients are the
same for all three kinds of operators, the representations depend on the
operators $M^{ab}$. After solving the
eigenvalue problem  for  invariants of 
subgroups, the representations can be presented as polynomials
of coordinates $\theta^a,$ or $ dx^a \wedge,$ $a =
0,1,2,3,5,..,14 $. The operators 
of spinorial character define the fundamental representations of
the group and the subgroups, while the operators of vectorial
character define the adjoint representations of the groups.
 We shall from now on, for the
sake of simplicity, refer to the polynomials of 
 Grassmann coordinates only. \\

We first analyze the space
of $2^d$ vectors for $d=14$ with respect to commuting operators
(Casimirs) of subgroups $SO(1,7)$ and $SO(6)$, so that 
 polynomials of $\theta^0, \theta^1, \theta^2, \theta^3,
\theta^5,\theta^6, \theta^7$ and $\theta^8$ are used to describe
states of the group 
SO(1,7) and then polynomials of $\theta^9, \theta^{10},
\theta^{11}, \theta^{12}, \theta^{13}$ and $\theta^{14}$
further to describe states of the group $SO(6)$. The group
$SO(1,13)$ has the rank equal to $r=7$, since it has $7 $
commuting operators (namely for example ${\cal S}^{01}, {\cal S}^{12},
{\cal S}^{35}, ...,{\cal S}^{13\;14} $), while the ranks of the
subgroups $SO(1,7)$ and 
$ SO(6)$ are accordingly $r=4$ and $r=3$, respectively. We may
further decide to arrange the basic states in the space of
polynomials of $\theta^0,...,\theta^8$ as eigenstates of $4 $
Casimirs of the subgroups $SO(1,3), SU(2),$ and $U(1)$ (the 
 first has  $r=2$, the second and the third have $r=1 $)  of the
group $SO(1,7)$, and the basic states in the space of polynomials of
$\theta^9,...,\theta^{14}$ as eigenstates of $r=3$ Casimirs of
subgroups $SU(3)$ and $U(1)$ ( with $r=2$ and $r=1$,
respectively) of the group $SO(6)$.  \\

We presented in Table I the eight Weyl spinors, two by two - 
one left ( $\tilde{\Gamma}^{(4)} = -1$) and one right (
$\tilde{\Gamma}^{(4)} = 1$) handed - 
connected by $\tilde{\gamma}^m, m=0,1,2,3$ into Weyl bi-spinors.
Half of vectors have Grassmann odd (odd products of
$\theta^m$ ) and half Grassmann even character.
The two four vectors of the same Grassmann character are
connected by the discrete time 
reversal operation $\theta^0 \rightarrow - \theta^0$ (
ref.(\cite{mankoc(99)})), while the two four vectors, which
differ in Grassmann character, are connected by the operation of
$\tilde{a}^a$.

According to Eqs.(\ref{taug},\ref{tauga}, \ref{taugb}), one can
express the 
generators of the subgroups $SU(2)$ and $U(1)$ of the group
$SO(1,7)$ in terms of the generators ${\cal S}^{ab}$.

We find (since the indices $0,1,2,3$ are reserved for the
subgroup $SO(1,3)$) 

\begin{eqnarray}
\tau^{31}: = \frac{1}{2} ( {\cal S}^{58} - {\cal S}^{67} ),\quad
\tau^{32}: = \frac{1}{2} ( {\cal S}^{57} + {\cal S}^{68} ),\quad
\tau^{33}: = \frac{1}{2} ( {\cal S}^{56} - {\cal S}^{78} ).
\label{su2w}
\end{eqnarray}
One also finds
\begin{equation}
\tau^{41}: = \frac{1}{2} ( {\cal S}^{56} + {\cal S}^{78} ).
\label{u1w}
\end{equation}

The algebra of  Eq.(\ref{taug}) follows\footnote[2]{Since the
operators $\tau^{Ai}$ have an even Grassmann character, the
generalized commutation relations agree with the usual
commutators, denoted by $[\;,\;]$.}
\begin{equation}
\{\tau^{3i}, \tau^{3j}\} = i \epsilon_{ijk} \tau^{3k}, \quad
\{\tau^{41}, \tau^{3i}\} = 0. 
\label{csu2u1w}
\end{equation}
One notices that $\tau^{51}: = \frac{1}{2} ( {\cal S}^{58} +
{\cal S}^{67} )$ and $\tau^{52}: = \frac{1}{2} ( {\cal S}^{57} -
{\cal S}^{68} )$ together with $ \tau^{41}$ form the algebra of
the group $SU(2)$ and that the generators of this group commute
with $\tau^{3i}$.

We present in Table II the eigenvectors of the operators
$\tilde{\tau}^{33}$ and $(\tilde{\tau}^3)^2 =
(\tilde{\tau}^{31})^2 + (\tilde{\tau}^{32})^2 
+(\tilde{\tau}^{33})^2 $, which are at the same time the
eigenvectors of 
$\tilde{\tau}^{41}$, for  spinors. We find, with respect to
the group $SU(2)$,   two doublets and four 
singlets  of an even and
another two doublets and four singlets of an odd Grassmann
character.

\begin{center}
\begin{tabular}{||c|c||c||c|c|c||}
\hline 
\hline
&&&&&\\
a & i &$<\theta|\Phi^a{ }_i>$& $\tilde{\tau}^{33}$& $
\tilde{\tau}^{41}$& Grassmann\\ 
&&&&&character\\
\hline
\hline
&&&&&\\
1 & 1 &$\frac{1}{2}(1-i\tilde{a}^5 \tilde{a}^6)(1+ i\tilde{a}^7
\tilde{a}^8)$ &$-\frac{1}{2}$ &$ 0$&\\
1 & 2 &$-\frac{1}{2}(\tilde{a}^5 + i\tilde{a}^6)(
\tilde{a}^7 - i \tilde{a}^8) $ &$\frac{1}{2} $&$ 0$&\\
&&&&&\\
2 & 1 &$\frac{1}{2}(1+i\tilde{a}^5 \tilde{a}^6)(1- i\tilde{a}^7
\tilde{a}^8)$ &$\frac{1}{2}$ &$ 0$&\\
2 & 2 &$-\frac{1}{2}(\tilde{a}^5 - i\tilde{a}^6)(
\tilde{a}^7 + i \tilde{a}^8) $ &$-\frac{1}{2} $&$ 0$&\\
&&&&&even\\
3 & 1 &$\frac{1}{2}(1+i\tilde{a}^5 \tilde{a}^6)(1+ i\tilde{a}^7
\tilde{a}^8)$ &0&$\frac{1}{2}$ &\\
4 & 1 &$\frac{1}{2}(\tilde{a}^5 + i\tilde{a}^6)(
\tilde{a}^7 + i \tilde{a}^8) $ &0&$\frac{1}{2} $&\\
5 & 1 &$\frac{1}{2}(1-i\tilde{a}^5 \tilde{a}^6)(1- i\tilde{a}^7
\tilde{a}^8)$ &0&$-\frac{1}{2}$ &\\
6 & 1 &$\frac{1}{2}(\tilde{a}^5 - i\tilde{a}^6)(
\tilde{a}^7 - i \tilde{a}^8) $ &0&$-\frac{1}{2} $&\\
&&&&&\\
\hline
\hline
&&&&&\\
7 & 1 &$\frac{1}{2}(1+i\tilde{a}^5 \tilde{a}^6)(\tilde{a}^7-i
\tilde{a}^8)$ &$\frac{1}{2}$ &$ 0$&\\
7 & 2 &$-\frac{1}{2}(\tilde{a}^5 - i\tilde{a}^6)(1+
\tilde{a}^7 \tilde{a}^8) $ &$-\frac{1}{2} $&$ 0$&\\
&&&&&\\
8 & 1 &$\frac{1}{2}(1-i\tilde{a}^5 \tilde{a}^6)(\tilde{a}^7+i
\tilde{a}^8)$ &$-\frac{1}{2}$ &$ 0$&\\
8 & 2 &$-\frac{1}{2}(\tilde{a}^5 + i\tilde{a}^6)(1-i
\tilde{a}^7 \tilde{a}^8) $ &$\frac{1}{2} $&$ 0$&\\
&&&&&odd\\
9 & 1 &$\frac{1}{2}(1-i\tilde{a}^5 \tilde{a}^6)(\tilde{a}^7-i
\tilde{a}^8)$ &0&$-\frac{1}{2}$ &\\
10 & 1 &$\frac{1}{2}(\tilde{a}^5 + i\tilde{a}^6)(1+
\tilde{a}^7 \tilde{a}^8) $ &0&$\frac{1}{2} $&\\
11 & 1 &$\frac{1}{2}(1+i\tilde{a}^5 \tilde{a}^6)(\tilde{a}^7+i
\tilde{a}^8)$ &0&$\frac{1}{2}$ &\\
12 & 1 &$\frac{1}{2}(\tilde{a}^5 - i\tilde{a}^6)(1-
\tilde{a}^7 \tilde{a}^8) $ &0&$-\frac{1}{2} $&\\
&&&&&\\
\hline
\hline
\end{tabular}
\end{center}

\noindent
Table II: The eigenstates of the operators $\tilde{\tau}^{33},
\tilde{\tau}^{41}$ are 
presented. We find two doublets and four singlets of an even
Grassmann character and two doublets and four singlets of an odd
Grassmann character. One sees that complex conjugation
transforms one doublet of either odd or even Grassmann character
into another of the same Grassmann character changing the signum
of  the
value of $\tilde{\tau}^{33}$, while it transforms one singlet into
another singlet
 of the same Grassmann character and of the opposite value of $
\tilde{\tau}^{41}$.  One can check that $\tilde{a}^h, \;\;
h \in (5,6,7,8)$, transforms the doublets of an even Grassmann
character into 
singlets of an odd Grassmann character.

One sees that $\tilde{\tau}^{5i},\;i = 1,2$, transform doublets into
singlets (which can easily be understood if taking into account
that $\tilde{\tau}^{5i}$ close together with $\tau^{41}$ the algebra 
of $SU(2)$ and that the two $SU(2)$ groups are isomorphic to the
group $SO(4)$).

One also sees the following very important property of 
representations of the 
group $SO(1,7) $: {\it If applying the
operators $\tilde{S}^{ab}$, 
$a,b = 0,1,2,3,5,6,7,8$ on the direct product of polynomials of
Table I and Table II, which forms the representations of the
group $SO(1,7)$, one finds that a 
multiplet of $SO(1,7)$ exists, which contains  left handed $SU(2)$ doublets
and right handed $SU(2)$ singlets.} It exists also another multiplet
which contains left handed $SU(2)$ singlets
and right handed $SU(2)$ doublets. It turns out that the operators
$\tilde{S}^{mh}$, with $m=0,1,2,3$ and $h=5,6,7,8$, although
having an even Grassmann 
character,  change the Grassmann character of that part of the
polynomials which belong to Table I and Table II,
respectively, keeping the 
Grassmann character of the products of the two types of
polynomials unchanged.  This can  be
understood if taking into account that  $\tilde{S}^{mh} =
-\frac{i}{2} \tilde{a}^m \tilde{a}^h$ and that the operator
$\tilde{a}^m$ changes the  polynomials  of an odd Grassmann character of
Table I, into an even polynomial, transforming a left handed
Weyl spinor of one family 
into a right handed Weyl spinor of another family, 
while $\tilde{a}^{h}$ changes simultaneously the $SU(2)$ 
 doublet of an even Grassmann character into a singlet of
an odd Grassmann character.

The symmetry, called the mirror symmetry, presented in this
approach, is not broken, as none of the symmetry is
broken. We only have arranged basic states to demonstrate 
 possible symmetries.

We  can  express   the
generators of  subgroups $SU(3)$ and $U(1)$ of the group
$SO(6)$ in terms of the generators ${\cal S}^{ab}$ (according to
Eq.(\ref{tauga})).

We find (since the indices $9,10,11,12,13,14$ are reserved for the
subgroup $SO(6)$) 
\begin{eqnarray}
\tau^{61}: = \frac{1}{2} ( {\cal S}^{9\;12} - {\cal S}^{10\;11} ),\quad
\tau^{62}: = \frac{1}{2} ( {\cal S}^{9\;11} + {\cal S}^{10\;12} ),\quad
\tau^{63}: = \frac{1}{2} ( {\cal S}^{9\;10} - {\cal S}^{11\;12} ),\quad
\nonumber
\\
\tau^{64}: = \frac{1}{2} ( {\cal S}^{9\;14} - {\cal S}^{10\;13} ),\quad
\tau^{65}: = \frac{1}{2} ( {\cal S}^{9\;13} + {\cal S}^{10\;14} ),\quad
\tau^{66}: = \frac{1}{2} ( {\cal S}^{11\;14} - {\cal S}^{12\;13}
),\quad 
\nonumber
\end{eqnarray}
\begin{eqnarray}
\tau^{67}: = \frac{1}{2} ( {\cal S}^{11\;13} + {\cal S}^{12\;14} ),\quad
\tau^{68}: = \frac{1}{2\sqrt{3}} ( {\cal S}^{9\;10} + {\cal
S}^{11\;12}  - 2{\cal S}^{13\;14}).\\
\label{su3c}
\end{eqnarray}

One  finds in addition 
\begin{equation}
\tau^{71}: = -\frac{1}{3}( {\cal S}^{9\;10} + {\cal S}^{11\;12}
+ {\cal S}^{13\;14} ).
\label{u1c}
\end{equation}

The algebra for the subgroups $SU(3)$ and $U(1)$ follows from
the algebra of the Lorentz group $SO(1,13)$
\begin{equation}
\{\tau^{6i}, \tau^{6j}\} = i f_{ijk} \tau^{6k}, \quad
\{\tau^{71}, \tau^{6i}\} = 0, {\rm \;\; for\;\; each \;\;i}. 
\label{csu3u1c}
\end{equation}
The coefficients $f_{ijk}$ are the structure constants of the group
$SU(3)$. 

We  can find the eigenvectors of the Casimirs of the groups
$SU(3)$ and $U(1)$  for spinors  as polynomials of
$\theta^h$, $h=9,...,14$. The eigenvectors, which are
polynomials of an 
even Grassmann character, can be found in
ref.(\cite{mankoc(92-97)}). We shall present here only not yet 
published (\cite{mankocbor(99)}) polynomials
of an odd Grassmann character.

\begin{center}
\begin{tabular}{||c|c||c||c|c|c||}
\hline
\hline 
&&&&&\\
a & i &$<\theta|\Phi^a{ }_i>$& $\tilde{\tau}^{63}$& $
\tilde{\tau}^{68}$& $ \tilde{\tau}^{71}$\\ 
&&&&&\\
\hline
\hline
&&&&&\\
1 & 1 & $\frac{1}{\sqrt{2^3}}(1+i\tilde{a}^{13} \tilde{a}^{14})(\tilde{a}^9
-i \tilde{a}^{10}) (1+i\tilde{a}^{11} \tilde{a}^{12}) $
& $\frac{1}{2}$ & $ \frac{1}{2\sqrt{3}} $ & $ \frac{1}{6}$ \\
1 & 2 &$\frac{1}{\sqrt{2^3}}(1+i\tilde{a}^{13}
\tilde{a}^{14})(1+i\tilde{a}^9 \tilde{a}^{10})
(\tilde{a}^{11}-i\tilde{a}^{12})$ 
&$-\frac{1}{2}$ &$ \frac{1}{2\sqrt{3}}$ & $\frac{1}{6}$ \\
1 & 3 &$-\frac{1}{\sqrt{2^3}}(\tilde{a}^{13}-i \tilde{a}^{14})(1+i\tilde{a}^9
 \tilde{a}^{10}) (1+i\tilde{a}^{11}\tilde{a}^{12})$
&$0$ &$ -\frac{1}{\sqrt{3}}$& $\frac{1}{6}$ \\
&&&&&\\
2 & 1 &$\frac{1}{\sqrt{2^3}}(1+i\tilde{a}^{13} \tilde{a}^{14})(1-i\tilde{a}^9
\tilde{a}^{10}) (\tilde{a}^{11}+i\tilde{a}^{12})$
&$\frac{1}{2}$ &$ \frac{1}{2\sqrt{3}}$ & $\frac{1}{6}$ \\
2 & 2 &$\frac{1}{\sqrt{2^3}}(1+i\tilde{a}^{13}
\tilde{a}^{14})(\tilde{a}^9+i \tilde{a}^{10})
(1-i\tilde{a}^{11}\tilde{a}^{12})$ 
&$-\frac{1}{2}$ &$ \frac{1}{2\sqrt{3}}$& $\frac{1}{6}$ \\
2 & 3 &$-\frac{1}{\sqrt{2^3}}(\tilde{a}^{13}-i \tilde{a}^{14})(\tilde{a}^9+i
 \tilde{a}^{10}) (\tilde{a}^{11}+i\tilde{a}^{12})$
&$0$ &$ -\frac{1}{\sqrt{3}}$& $\frac{1}{6}$ \\
&&&&&\\
3 & 1 &$\frac{1}{\sqrt{2^3}}(\tilde{a}^{13}+i\tilde{a}^{14})(\tilde{a}^9
-i \tilde{a}^{10}) (\tilde{a}^{11}+i\tilde{a}^{12})$
&$\frac{1}{2}$ &$ \frac{1}{2\sqrt{3}}$& $\frac{1}{6}$ \\
3 & 2 &$\frac{1}{\sqrt{2^3}}(\tilde{a}^{13}+i
\tilde{a}^{14})(1+i\tilde{a}^9 \tilde{a}^{10})
(1-i\tilde{a}^{11}\tilde{a}^{12})$ 
&$-\frac{1}{2}$ &$ \frac{1}{2\sqrt{3}}$& $\frac{1}{6}$ \\
3 & 3 &$-\frac{1}{\sqrt{2^3}}(1-i\tilde{a}^{13} \tilde{a}^{14})(1+i\tilde{a}^9
 \tilde{a}^{10}) (\tilde{a}^{11}+i\tilde{a}^{12})$
&$0$ &$ -\frac{1}{\sqrt{3}}$& $\frac{1}{6}$ \\
&&&&&\\
4 & 1 &$\frac{1}{\sqrt{2^3}}(\tilde{a}^{13}+i\tilde{a}^{14})(1-i\tilde{a}^9
\tilde{a}^{10}) (1+i\tilde{a}^{11}\tilde{a}^{12})$
&$\frac{1}{2}$ &$ \frac{1}{2\sqrt{3}}$& $\frac{1}{6}$ \\
4 & 2 &$\frac{1}{\sqrt{2^3}}(\tilde{a}^{13}+i
\tilde{a}^{14})(\tilde{a}^9+i \tilde{a}^{10})
(\tilde{a}^{11}-i\tilde{a}^{12})$ 
&$-\frac{1}{2}$ &$ \frac{1}{2\sqrt{3}}$ & $\frac{1}{6}$ \\
4 & 3 &$-\frac{1}{\sqrt{2^3}}(1-i\tilde{a}^{13} \tilde{a}^{14})(\tilde{a}^9
+i \tilde{a}^{10}) (1+i\tilde{a}^{11}\tilde{a}^{12})$
&$0$ &$ -\frac{1}{\sqrt{3}}$ & $\frac{1}{6}$ \\
&&&&&\\
\hline
\hline
&&&&&\\
5 & 1 & $\frac{1}{\sqrt{2^3}}(1-i\tilde{a}^{13} \tilde{a}^{14})(\tilde{a}^9
+i \tilde{a}^{10}) (1-i\tilde{a}^{11} \tilde{a}^{12}) $
& $-\frac{1}{2}$ & $ -\frac{1}{2\sqrt{3}} $ & $ -\frac{1}{6}$ \\
5 & 2 &$\frac{1}{\sqrt{2^3}}(1-i\tilde{a}^{13}
\tilde{a}^{14})(1-i\tilde{a}^9 \tilde{a}^{10})
(\tilde{a}^{11}+i\tilde{a}^{12})$ 
&$\frac{1}{2}$ &$- \frac{1}{2\sqrt{3}}$ & $-\frac{1}{6}$ \\
5 & 3 &$-\frac{1}{\sqrt{2^3}}(\tilde{a}^{13}+i \tilde{a}^{14})(1-i\tilde{a}^9
 \tilde{a}^{10}) (1-i\tilde{a}^{11}\tilde{a}^{12})$
&$0$ &$ \frac{1}{\sqrt{3}}$& $-\frac{1}{6}$ \\
&&&&&\\
6 & 1 &$\frac{1}{\sqrt{2^3}}(1-i\tilde{a}^{13} \tilde{a}^{14})(1+i\tilde{a}^9
\tilde{a}^{10}) (\tilde{a}^{11}-i\tilde{a}^{12})$
&$-\frac{1}{2}$ &$ -\frac{1}{2\sqrt{3}}$ & $-\frac{1}{6}$ \\
6 & 2 &$\frac{1}{\sqrt{2^3}}(1-i\tilde{a}^{13}
\tilde{a}^{14})(\tilde{a}^9-i \tilde{a}^{10})
(1+i\tilde{a}^{11}\tilde{a}^{12})$ 
&$\frac{1}{2}$ &$ -\frac{1}{2\sqrt{3}}$& $-\frac{1}{6}$ \\
6 & 3 &$-\frac{1}{\sqrt{2^3}}(\tilde{a}^{13}+i \tilde{a}^{14})(\tilde{a}^9-i
 \tilde{a}^{10}) (\tilde{a}^{11}-i\tilde{a}^{12})$
&$0$ &$ \frac{1}{\sqrt{3}}$& $-\frac{1}{6}$ \\
&&&&&\\
7 & 1 &$\frac{1}{\sqrt{2^3}}(\tilde{a}^{13}-i\tilde{a}^{14})(\tilde{a}^9
+i \tilde{a}^{10}) (\tilde{a}^{11}-i\tilde{a}^{12})$
&$-\frac{1}{2}$ &$ -\frac{1}{2\sqrt{3}}$& $-\frac{1}{6}$ \\
7 & 2 &$\frac{1}{\sqrt{2^3}}(\tilde{a}^{13}-i
\tilde{a}^{14})(1-i\tilde{a}^9 \tilde{a}^{10})
(1+i\tilde{a}^{11}\tilde{a}^{12})$ 
&$\frac{1}{2}$ &$ -\frac{1}{2\sqrt{3}}$& $-\frac{1}{6}$ \\
7 & 3 &$-\frac{1}{\sqrt{2^3}}(1+i\tilde{a}^{13} \tilde{a}^{14})(1-+i\tilde{a}^9
 \tilde{a}^{10}) (\tilde{a}^{11}-i\tilde{a}^{12})$
&$0$ &$ \frac{1}{\sqrt{3}}$& $-\frac{1}{6}$ \\
&&&&&\\
8 & 1 &$\frac{1}{\sqrt{2^3}}(\tilde{a}^{13}-i\tilde{a}^{14})(1+i\tilde{a}^9
\tilde{a}^{10}) (1-i\tilde{a}^{11}\tilde{a}^{12})$
&$-\frac{1}{2}$ &$ -\frac{1}{2\sqrt{3}}$& $-\frac{1}{6}$ \\
8 & 2 &$\frac{1}{\sqrt{2^3}}(\tilde{a}^{13}-i
\tilde{a}^{14})(\tilde{a}^9-i \tilde{a}^{10})
(\tilde{a}^{11}+i\tilde{a}^{12})$ 
&$\frac{1}{2}$ &$ -\frac{1}{2\sqrt{3}}$ & $-\frac{1}{6}$ \\
8 & 3 &$-\frac{1}{\sqrt{2^3}}(1+i\tilde{a}^{13} \tilde{a}^{14})(\tilde{a}^9
-i \tilde{a}^{10}) (1-i\tilde{a}^{11}\tilde{a}^{12})$
&$0$ &$ \frac{1}{\sqrt{3}}$ & $-\frac{1}{6}$ \\
&&&&&\\
\hline
\hline
&&&&&\\
9 & 1 &$\frac{1}{\sqrt{2^3}}(1+i\tilde{a}^{13} \tilde{a}^{14})(\tilde{a}^9
+i \tilde{a}^{10}) (1+i\tilde{a}^{11}\tilde{a}^{12})$
&$0$ &$ 0$ & $\frac{1}{2}$ \\
10 & 1 &$\frac{1}{\sqrt{2^3}}(1+i\tilde{a}^{13}
\tilde{a}^{14})(1+i \tilde{a}^9 \tilde{a}^{10})
(\tilde{a}^{11}+i\tilde{a}^{12})$ 
&$0$ &$ 0 $ & $\frac{1}{2}$ \\
11 & 1 &$\frac{1}{\sqrt{2^3}}(\tilde{a}^{13}+i \tilde{a}^{14})(1+i\tilde{a}^9
 \tilde{a}^{10}) (1+i\tilde{a}^{11}\tilde{a}^{12})$
&$0$ &$ 0$ & $\frac{1}{2}$ \\
12 & 1 &$\frac{1}{\sqrt{2^3}}(\tilde{a}^{13}+i \tilde{a}^{14})(\tilde{a}^9
+i \tilde{a}^{10}) (\tilde{a}^{11}+i\tilde{a}^{12})$
&$0$ &$ 0$ & $\frac{1}{2}$ \\
13 & 1 &$\frac{1}{\sqrt{2^3}}(1-i\tilde{a}^{13}
\tilde{a}^{14})(\tilde{a}^9-i \tilde{a}^{10})
(1-i\tilde{a}^{11}\tilde{a}^{12})$ 
&$0$ &$ 0$ & $-\frac{1}{2}$ \\
14 & 1 &$\frac{1}{\sqrt{2^3}}(1-i\tilde{a}^{13} \tilde{a}^{14})(1-i\tilde{a}^9
 \tilde{a}^{10}) (\tilde{a}^{11}-i\tilde{a}^{12})$
&$0$ &$ 0$ & $-\frac{1}{2}$ \\
15 & 1 &$\frac{1}{\sqrt{2^3}}(\tilde{a}^{13}-i
\tilde{a}^{14})(1-i\tilde{a}^9 \tilde{a}^{10})
(1-i\tilde{a}^{11}\tilde{a}^{12})$ 
&$0$ &$ 0$ & $-\frac{1}{2}$ \\
16 & 1 &$\frac{1}{\sqrt{2^3}}(\tilde{a}^{13}-i \tilde{a}^{14})(\tilde{a}^9-i
 \tilde{a}^{10}) (\tilde{a}^{11}-i\tilde{a}^{12})$
&$0$ &$ 0$ & $-\frac{1}{2}$ \\
&&&&&\\
\hline
\hline
\end{tabular}
\end{center}

\noindent
Table III: The eigenstates of the operators $\tilde{\tau}^{63},
\tilde{\tau}^{68}, \tilde{\tau}^{71}$ are presented for odd
Grassmann character polynomials. We find four triplets, four
antitriplets and eight singlets.
 One sees that complex conjugation
transforms one triplet into antitriplet, while
$\tilde{\tau}^{8i}$ transform triplets into antitriplets or singlets.

One finds four triplets and four antitriplets as well as eight
singlets. Besides the eigenvalues of the commuting operators
$\tilde{\tau}^{63}$ 
and $\tilde{\tau}^{68}$ of the group $SU(3)$ also the eigenvalue
of $\tilde{\tau}^{71}$ forming $U(1)$, is presented.
The operators $\tilde{\tau}^{81}: = \frac{1}{2} (
\tilde{S}^{9\;12} + \tilde{S}^{10\;11} ),\quad 
\tau^{82}: = \frac{1}{2} ( \tilde{S}^{9\;11} - \tilde{S}^{10\;12} ),\quad
\tau^{83}: = \frac{1}{2} ( \tilde{S}^{9\;14} + \tilde{S}^{10\;13} ),\quad
\tau^{84}: = \frac{1}{2} ( {\cal S}^{9\;13} - {\cal S}^{10\;14} ),\quad
\tau^{85}: = \frac{1}{2} ( \tilde{S}^{11\;14} + \tilde{S}^{12\;13}),\quad 
\tau^{86}: = \frac{1}{2} ( \tilde{S}^{11\;13} - \tilde{S}^{12\;14}),\quad 
 $ which  transform triplets of the group $SU(3)$ into
antitriplets and singlets with respect to the group $SU(3)$.

The spinorial representations of the group $SO(1,13)$ are the 
direct product of polynomials of Table I, Table II and
Table III.

We can find all the members of a spinorial multiplet of the
group $SO(1,13)$ by 
applying $\tilde{S}^{ab}$ on any initial Grassmann odd product of 
polynomials, if one polynomial is taken from Table I, another from Table
II and the third from Table III. In the same multiplet there are
triplets, singlets and 
antitriplets with respect to $SU(3)$, which are doublets or singlets
with respect to $SU(2)$, and are left and right handed with
respect to $SO(1,3)$. 

We can arrange in the same sense also eigenstates of operators
of vectorial character, with bosonic character. In this paper we
shall not do that.

\subsection{Dynamical arrangement of representations of $SO(1,13)$
with respect to subgroups $SO(1,7)$ and $SO(6)$. \label{dar}}

To see  how  Yang-Mills fields enter into the theory,
we shall rewrite the Weyl-like equation in the presence of the
gravitational field (\ref{cong}) in terms of
components of fields which determine  
gravitation in the four dimensional subspace and of those 
which determine  gravitation in higher dimensions, assuming
that the coordinates of ordinary space with indices higher than
four stay compacted to unmeasurable small dimensions (or can not
at all be noticed for some other reason). 
Since  Grassmann space only manifests itself through  average
values of observables, compactification of a part of 
Grassmann space has no meaning.  However, since
parameters of  Lorentz transformations in a freely falling
coordinate system for both spaces have
to be the same, no transformations to the fifth or
higher coordinates may occur 
at measurable energies. Therefore, at low energies, the four
dimensional subspace 
of  Grassmann space with the generators defining the Lorentz
group $ SO(1,3)$ is (almost) decomposed from the rest of the 
Grassmann space with the generators forming the (compact) group
$ SO(d-4) $, because of the decomposition of  ordinary
space. This is valid on the classical level only.

According to the previous subsection, the break of symmetry
of $SO(1,13)$
should, however, appears in steps,  first  through
$SO(1,7)\times SO(6)$ and later to the final symmetry, which is
needed in the Standard model for massless particles.

We shall comment on possible ways of spontaneously broken
symmetries by studying the Weyl equation in the presence of
gravitational fields in d dimensions for massless particles
(Eqs.(\ref{cmg}, \ref{cong}))
\begin{equation}
\tilde{\gamma}^{a} p_{0a} = 0, \quad p_{0a} = f^{\mu}{ }_a 
p_{0\mu}, \quad p_{0\mu} = p_{\mu} - \frac{1}{2}
\tilde{S}^{ab}\omega_{ab\mu}.  
\label{dgebb}
\end{equation}
\subsubsection{Standard model case. \label{smodc}}

To make discussions more transparent we shall first comment on
 the well known case  of the Standard model. Before the break of the
symmetry $SU(3) \times SU(2) \times U(1)$ into $SU(3) \times
U(1)$, the canonical momentum $p_{0\alpha}$ ($\; \alpha = 
0,1,2,3$ and $d=4$)  includes the gauge fields, connected with the
groups $SU(3)$, $SU(2)$ and $U(1)$. We shall pay attention
on only 
the groups $SU(2)$ and $U(1)$, which are involved in the break
of symmetry 
\begin{equation}
p_{0\alpha} = p_{\alpha} - g \tau^{i}A^{i}{ }_{\alpha} - g'Y B_{\alpha},  
\label{smp}
\end{equation} 
where $g$ and $g'$ are the two coupling constants.
Introducing $\tau^{\pm} = \tau^1 \pm i \tau^2$, the
superposition follows $ A^{\pm}{
}_{\alpha} = A^1{ }_{\alpha} \mp i \; A^2{ }_{\alpha} $. If defining $ A^3{
}_{\alpha} = \frac{g/g'}{\sqrt{1 + (g/g')^2}} Z_{\alpha} + \frac{1}{\sqrt{1
+ (g/g')^2}} A_{\alpha}$ and $B_{\alpha} = -\frac{1}{\sqrt{1 +
(g/g')^2}} Z_{\alpha}  +
\frac{g/g'}{\sqrt{1 + (g/g')^2}} A_{\alpha}$, so that the
transformation is orthonormalized, one can easily rewrite 
Eq.(\ref{smp}) as follows

\begin{equation}
p_{0\alpha} = p_{\alpha} - \frac{g}{2} (\tau^{+}A^{+}{
}_{\alpha} + \tau^{-}A^{-}{ }_{\alpha}) + \frac{gg'}{\sqrt{g^2 +
g'^2}} Q A_{\alpha} + \frac{g^2}{\sqrt{g^2 + g'^2}} Q' Z_{\alpha}. 
\label{smpt}
\end{equation}
with
\begin{equation}
Q = \tau^3 + Y, \quad Q' = \tau^3 - (\frac{g'}{g})^2 Y.
\label{qq'}
\end{equation} 
 
In the Standard model $<Q>$ is conserved quantity and $<Q'>$ is
not, due to the fact that $<Q>$ is zero for the Higgs
fields in the ground state, while $<Q'>$ is nonzero ( 
$<Q'> = - \frac{1}{2} ( 1 + (\frac{g'}{g})^2)$). 

We further see that in the
case that $g = g'$, it follows that $Q = \tau^3 + Y$ and $Q' =
\tau^3 - Y$.  If no symmetry is spontaneusly  broken, that is if no
Higgs breaks symmetry by making a choice for his ground state
symmetry, the only thing which has been done by introducing
linear superpositions of fileds, is the 
rearrangement of fields, which  always can be done without any
consequence, except that it may help to better see the
symmetries. 

Spontaneusly broken symmetries couse the nonconservation of
quantum numbers, as well as massive clusters of fields.

\subsubsection{Spin connections and gauge fields leading to the
Standard model. \label{scgf}}

We shall rewrite the canonical momentum of Eq.(\ref{dgebb})
to manifest possible ways of breaking symmetries of $SO(1,13)$
down to the symmetries of the Standard model. We first write 

\begin{equation}
\tilde{\gamma}^{a} p_{0a} = 0 = \tilde{\gamma}^{a} f^{\mu}{ }_a
p_{0\mu} = (\tilde{\gamma}^{m} f^{\alpha}{ }_m  +
\tilde{\gamma}^{h} f^{\alpha}{ }_h )
p_{0\alpha}  + (\tilde{\gamma}^{m} f^{\sigma}{ }_m  +
\tilde{\gamma}^{h} f^{\sigma}{ }_h  ) p_{0\sigma},
\label{dgex}
\end{equation}
with $\alpha, m \in \{0,1,2,3 \} $ and $\sigma, h \in \{5,...,14
\} $ to separate the $d=4$ dimensional subspace out of $d = 14$
dimensional space. We may further rearrange the canonical
momentum $p_{0\mu}$
\begin{equation}
p_{0\mu} = p_{\mu} - \frac{1}{2} \tilde{S}^{h_1 h_2}\;
\omega_{h_1 h_2 \mu}- \frac{1}{2} \tilde{S}^{k_1 k_2}\;
\omega_{k_1 k_2 \mu} - \frac{1}{2} \tilde{S}^{h_1 k_1}\;
\omega_{h_1 k_1 \mu},
\label{p0a}
\end{equation}
with $h_i \in \{0,1,..,8 \}$ and $k_i \in \{9,...,14 \}$
so that $\tilde{S}^{h_1h_2}$ define the algebra of the subgroup
$SO(1,7)$, while $\tilde{S}^{h_1h_2}$ define the algebra of the
subgroup $SO(6)$. The generators $\tilde{S}^{h_1k_1}$ rotate 
states of a multiplet of the group $SO(1,13)$ into each other.  

Taking into account subsection \ref{ac}  we may
rewrite the generators $\tilde{S}^{ab}$ in terms of the
corresponding generators of subgroups $\tilde{\tau}^{Ai}$ and
accordingly, similarly to the Standard model case, introduce
new fields (see subsection \ref{smodc}), which are 
superpositions of the old ones
\begin{eqnarray}
g \;A^{31}{ }_{\mu} = \frac{1}{2} (\omega_{58\mu} - \omega_{67\mu}), \quad
g \;A^{32}{ }_{\mu} = \frac{1}{2} (\omega_{57\mu} + \omega_{68\mu}), \quad
g \;A^{33}{ }_{\mu} = \frac{1}{2} (\omega_{56\mu} - \omega_{78\mu}), \quad
\label{su2a}
\\ 
g \;A^{41}{ }_{\mu} = \frac{1}{2} (\omega_{56\mu} + \omega_{78\mu}), \quad
\nonumber
\\
g \;A^{51}{ }_{\mu} = \frac{1}{2} (\omega_{58\mu} + \omega_{67\mu}), \quad
g \;A^{52}{ }_{\mu} = \frac{1}{2} (\omega_{57\mu} - \omega_{68\mu}). \quad
\label{su2u1af}
\end{eqnarray} 
It follows then 
\begin{equation}
\frac{1}{2} \tilde{S}^{h_1h_2}\; \omega_{h_1 h_2 \mu} =
g \; \tilde{\tau}^{Ai} A^{Ai}{ }_{\mu},
\label{su2u1a}
\end{equation}
where for $A=3$,  $i = 1,2,3$, for $A=4$,  $i = 1$ and for $A=5$
$i=1,2$. Accordingly, the fields $A^{Ai}_{}\mu$ are the gauge
fields of the group $SU(2)$, if $A=3$ and of $U(1)$ if $A=4$. 
Since $\tilde{\tau}^{41}$ and $\tilde{\tau}^{5i}$ form the group
$SU(2)$ as well, the corresponding fields could be the gauge
fields of this group. The break of symmetry should make a choice
between  the gauge groups $U(1)$ and $SU(2)$.

We leave the notation for spin connection fields in the case that
$h_i \in \{ 0,1,2,3\}$ unchanged. We also leave unchanged the 
spin connection fields for the case, that $h_1 = 0,1,2,3$ and
$h_2 = 5,6,7,8$ as well as for the case, that $h_1 \in
\{0,1...,8 \}$ and $k_1 \in \{9,..,14 \}$, while we  arrange 
terms with $k_i \in \{8,...,14\} $ to demonstrate the symmetry
$SU(3)$ and $U(1)$
\begin{eqnarray}
g \;A^{61}{ }_{\mu} &=& \frac{1}{2} (\omega_{9\;12\mu} - \omega_{10\;11\mu}), \quad
g \;A^{62}{ }_{\mu} = \frac{1}{2} (\omega_{9\;11\mu} + \omega_{10\;12\mu}), \quad
g \;A^{63}{ }_{\mu} = \frac{1}{2} (\omega_{9\;10\mu} - \omega_{11\;12\mu}), \quad 
\nonumber
\\
g \;A^{64}{ }_{\mu} &=& \frac{1}{2} (\omega_{9\;14\mu} - \omega_{10\;13\mu}), \quad
g \;A^{65}{ }_{\mu} = \frac{1}{2} (\omega_{9\;13\mu} + \omega_{10\;14\mu}), \quad
g \;A^{66}{ }_{\mu} = \frac{1}{2} (\omega_{11\;14\mu} - \omega_{12\;13\mu}), \quad 
\nonumber
\\
g \;A^{67}{ }_{\mu} &=& \frac{1}{2} (\omega_{11\;13\mu} + \omega_{12\;14\mu}), \quad
g \;A^{68}{ }_{\mu} = \frac{1}{2\sqrt{3}} (\omega_{9\;10\mu} + \omega_{11\;12\mu} - 
2  \omega_{13\;14\mu}),\quad
\label{su3a}
\\
g \;A^{71}{ }_{\mu} &=& - \frac{1}{2} (\omega_{9\;10\mu} + \omega_{11\;12\mu} + 
\omega_{13\;14\mu}).\quad
\label{su3u1a}
\end{eqnarray}
We may accordingly define fields $
g \;A^{81}{ }_{\mu} = \frac{1}{2} (\omega_{9\;12\mu} + \omega_{10\;11\mu}), \quad 
g \;A^{82}{ }_{\mu} = \frac{1}{2} (\omega_{9\;11\mu} -
\omega_{10\;12\mu}), \quad $ $
g \;A^{83}{ }_{\mu} = \frac{1}{2} (\omega_{9\;14\mu} + \omega_{10\;13\mu}), \quad 
g \;A^{84}{ }_{\mu} = \frac{1}{2} (\omega_{9\;13\mu} - \omega_{10\;14\mu}), \quad
g \;A^{85}{ }_{\mu} = \frac{1}{2} (\omega_{11\;14\mu} +
\omega_{12\;13\mu}), \quad $ $
g \;A^{86}{ }_{\mu} = \frac{1}{2} (\omega_{11\;13\mu} - \omega_{12\;14\mu}) \quad 
 $, so that it follows
\begin{equation}
\frac{1}{2} \tilde{S}^{k_1k_2}\; \omega_{k_1 k_2 \mu} =
g \; \tilde{\tau}^{Ai} A^{Ai}{ }_{\mu},
\label{su3u1gf}
\end{equation}
with $ A = 6,7,8$. While $A^{6i}{}_{\mu},\; i \in \{1,..,8 \}$, form the
gauge field of the group $SU(3)$ and $A^{71}{ }_{\mu}$
corresponds to the gauge group $U(1)$, terms $g
\tilde{\tau}^{7i}\; A^{7i}{ }_{\mu}$ transform $SU(3)$ triplets 
into singlets and antitriplets. Again, without additional
requirements, all the coupling constants  $g$ are equal.
To be in agreement with what the Standard model needs as an
input, we further  rearrange the gauge fields belonging to
the two $U(1)$ fields, one coming from the subgroup $SO(1,7)$
the other from the subgroup $SO(6)$.
We therefore define
\begin{equation}
Y_1 =  (\tau^{41} + \tau^{71}),\quad 
Y_2 = -  (\tau^{41} - \tau^{71})\quad 
\label{y1y2}
\end{equation}
and accordingly similarly to the Standard model case of
subsection \ref{smodc}
\begin{equation}
A^{1}{ }_{\mu} = \frac{1}{2}  (A^{41}{ }_{\mu}
+ A^{71}{ }_{\mu}), \quad 
A^{2}{ }_{\mu} = -\frac{1}{2}  (A^{41}{ }_{\mu}
- A^{71}{ }_{\mu}). \quad 
\label{y1y2a}
\end{equation}

The rearrangement of fields demonstrates all the symmetries of the
massless particles of the Standard model and more. 

Taking into account Tables I, II and III one finds for the
quantum numbers of spinors, which belong to a multiplet of
$SO(1,7)$ with left handed $SU(2)$ doublets and right handed
$SU(2)$ singlets and which are triplets or singlets with respect
to $SU(3)$, the ones, presented on Table IV. We use the
names of the Standard model to denote triplets and singlets with
respect to $SU(3)$ and $SU(2)$. 
\begin{center}
\begin{tabular}{||c|c|cccccc|cccccc||}
\hline
\hline
&&&&&&&&&&&&\\
  & & 
   \multicolumn{6}{c|}{SU(2) doublets} & 
   \multicolumn{6}{c|}{SU(2) singlets}\\ 
  & & $\tilde{\tau}^{33}$& $
\tilde{\tau}^{41} $& 
$\tilde{\tau}^{71} $&$\tilde{Y}_1$&$\tilde{Y}_2$ &
$\tilde{\Gamma}^{(4)}$ & $\tilde{\tau}^{33}$& 
$\tilde{\tau}^{41}$ & $\tilde{\tau}^{71}$ & $\tilde{Y}_1 $& $\tilde{Y}_2$&
$\tilde{\Gamma}^{(4)}$ \\ \hline
&&&&&&&&&&&&\\
SU(3) triplets & & &&&&&&&&&& \\
$\tilde{\tau}^{6\;3}\;\;$  =  & 
$u_i$& 
1/2 & 0 & 1/6 & 1/6 & 1/6 & - 1 & 0 & 1/2 & 1/6 & 2/3 & -1/3 & 1
\\ 
 ( $ \frac{1}{2},\;\;$  $ -\frac{1}{2},\;\;$
 $ 0\;$  )&&&&&&&&&&&\\
$\tilde{\tau}^{6\;8} $  = & 
$d_i$ & -1/2 & 0 & 1/6 & 1/6 & 1/6 & -1 & 0 & -1/2 & 1/6 & -1/3
& 2/3 & 1 \\
 ( $\frac{1}{2 \sqrt{3}},$
$\frac{1}{2 \sqrt{3}},$  $-\frac{1}{ \sqrt{3}}$  )&&&&&&&&&&&\\  \hline 
&&&&&&&&&&&&\\
SU(3) singlets & & &&&&&&&&&& \\
$\tilde{\tau}^{6\; 3} =  0$ & $\nu_i$ & 1/2 & 0 & -1/2 & 
-1/2 & -1/2 & -1 & 0 & 1/2 & -1/2 & 0 & -1 & 1\\
$\tilde{\tau}^{6\; 8} = 0$ & 
$e_i$ & -1/2 & 0 & -1/2 & 
-1 & -1 & -1 & 0 & -1/2 & -1/2 & -1 & 0 & 1\\ \hline
\hline
\end{tabular}
\end{center}

\noindent
Table IV: Expectation values for generators $\tilde{\tau}^{63}$
and $\tilde{\tau}^{68}$ of the group $SU(3)$ and the generator
$\tilde{\tau}^{71}$ of the group $U(1)$, the two groups are
 subgroups of the group $SO(6)$, and of generators o
$\tilde{\tau}^{33}$ of the group $SU(2)$ , $\tilde{\tau}^{41}$
of the group $U(1)$ and $\tilde{\Gamma}^{(4)}$ of the group
$SO(1,3)$, the three groups are subgroups of the group $SO(1,7)$
for the multiplet (with respect to $SO(1,7)$), which contains
left handed ($<\Gamma^{(4)}> = -1$) $SU(2)$ doublets and right
handed ($<\Gamma^{(4)}> = 1$) $SU(2)$ singlets. In addition, 
values for $\tilde{Y}_1$ and $\tilde{Y}_2$ are also presented.
Index $i$ of $u_i, d_i, \nu_i $ and $e_i$ runs over four
families presented in Table I.

We see that, besides $\tilde{Y}_2$, this are just the quantum
numbers needed for massless fermions of the Standard model. The
value for the additional hyper charge $\tilde{Y}_2$ is nonzero
for the right 
handed neutrinos, as well as for other states, except right
handed electrons.

Since no symmetry is broken yet, all the gauge fields are of the
same strength. To come to the symmetries of massless fields of
the Standard model,  surplus symmetries
should be broken so that all the coupling constants connected with
the fields $\omega_{ab\mu}$ which do not determine the fields
$A^{Ai}_{\mu}$, $A=3,6$ (Eqs.(\ref{su2a},\ref{su3a})) and $A^1_{\mu}$
(Eq.(\ref{y1y2a})) should be small and yet the coupling
constants of these three fields should not be equal. Accordingly
also the operator $\tilde{Y}_2$ could, similarly to the case of 
Eq.(\ref{qq'}), depend on the coupling constants. 

The mirror symmetry should be broken so that multiplets
of $SO(1,7)$ with right handed $SU(2)$ doublets and left handed
$SU(2)$ singlets become very massive.  All the surplus multiplets,
either bosonic or fermionic should become of large enough masses
not to be measurable yet.
 
The proposed approach predicts four rather than three families
of fermions.

Although in this paper, we shall not discuss possible ways of
appearance of spontaneously broken 
symmetries, bringing the symmetries of the group $SO(1,13)$
down to symmetries of the Standard model, we still would like to
know, whether there are terms in the Weyl equation
(Eq.\ref{dgex}) which may behave like the Yukawa couplings. We
see that indeed the term $\tilde{\gamma}^{h}f^{\sigma}{ }_h p_{0
\sigma}$, with $h \in \{ 5,6,7,8 \}$ and $\sigma \in \{5,6,..
\}$ really may, if operating on a right handed $SU(2)$ singlet
transform it to a left handed $SU(2)$ doublet. We also can find
among scalars the terms with quantum numbers of Higgs bosons
(which are $SU(2)$ doublets with respect to operators of the
vectorial character.)
All this is in preparation and not yet finished or fully understood.

\section{Concluding remarks. \label{cr}}

In this paper, we demonstrated that if assuming that the space has $d$
commuting and $d$ anticommuting coordinates, then, for $d\ge 14$, all
spins in $d$ dimensions, described in the vector space spanned
over the space of anticommuting
coordinates, demonstrate in four dimensional subspace
as spins and all charges, unifying spins and charges of fermions
and bosons independently, although the supersymmetry, which
means the same number of fermions and bosons, is a manifesting
symmetry.  The anticommuting coordinates can be represented by
either Grassmann coordinates or by the K\" ahler differential
forms. 

We demonstrated that either our approach or the approach of
differential forms suggest four families of quarks and leptons,
rather than three.

We have shown that starting (in any of the two approaches) with
the Lorentz symmetry in the tangent space in $d\ge 14$, spins
degrees of freedom ( described by dynamics in the space of
anticommuting coordinates) manifests in four dimensional subspace
as spins and colour, weak and hyper charges, with one additional
hyper charge, in a way that only left handed weak charge
doublets together with right handed weak charge singlets appear,
if the symmetry is spontaneously broken from $SO(1,13)$ first to
$SO(1,7)$ and $SO(6)$, so that a multiplet of $SO(1,7)$ with
only left handed $SU(2)$ doublets and right handed $SU(2)$
singlets survive, while the mirror symmetry is broken, and then
to $SO(1,3), SU(2), SU(3) $ and $U(1).$

We have demonstrated that the gravity in D dimensions manifests
as ordinary gravity and all gauge fields in four dimensional
subspace, after the break of symmetry and the accordingly
changed coupling constant.
We also have shown that there are terms in the Weyl equations,
which in four dimensional subspace manifest as yukawa couplings.

The two approaches, the K\" ahler one after the generalization,
which we have been suggested, and our, lead to the same results.

\section*{Acknowledgment } This work was supported by Ministry of
Science and Technology of Slovenia. The author would like to
acknowledge the work done together with Anamarija Bor\v stnik,
which is the breaking of the SO(1,13) symmetry.



\end{document}